\documentclass[usenatbib]{mn2e}

\usepackage{graphicx}
\usepackage{graphics}
\usepackage{rotating}
\usepackage{color}
\usepackage{amsmath}

\bibpunct{(}{)}{;}{a}{}{,}

\begin{document}

\def\HI{H{\sc i} }
\def\HII{H{\sc ii} }
\def\Ha{{\rm H}\alpha }
\def\Msun{{\mathrm M}_{\odot} }
\def\kms{\,km\,s^{-1}}
\def\Msunrm{{\mathrm M}_{\odot} }
\def\Lsunrm{{\mathrm L}_{\odot} }

\title[Chemo-dynamical evolution of tidal dwarf galaxies. II.]{Chemo-dynamical evolution of tidal dwarf galaxies. \\ II. The long-term evolution and influence of a tidal field}

\author[S. Ploeckinger et al.]
	{S.~Ploeckinger$^1$\thanks{email: sylvia.ploeckinger@univie.ac.at},  S.~Recchi$^1$,  G.~Hensler$^1$,  P.~Kroupa$^2$ \\
         $^1$University of Vienna, Department of Astrophysics, T\"urkenschanzstr. 17, 1180 Vienna, Austria  \\
	$^2$ Helmholtz-Institut f\"ur Strahlen- und Kernphysik, Nussallee 14-16, 53115 Bonn, Germany}

\maketitle

\begin{abstract}
In a series of papers, we present detailed chemo-dynamical simulations of tidal dwarf galaxies (TDGs). After the first paper, where we focused on the very early evolution, we present in this work simulations on the long-term evolution of TDGs, ranging from their formation to an age of 3 Gyr. 
Dark-matter free TDGs may constitute a significant component of the dwarf galaxy (DG) population.  But it remains to be demonstrated  that TDGs can survive their formation phase given stellar feedback processes, the time-variable tidal field of the post-encounter host galaxy and its dark matter halo and ram-pressure wind from the gaseous halo of the host. For robust results the maximally damaging feedback by a fully populated invariant stellar IMF in each star cluster is assumed, such that fractions of massive stars contribute during phases of low star-formation rates. The model galaxies are studied in terms of their star-formation history, chemical enrichment and rotational curves. All models evolve into a self-regulated long-term equilibrium star-formation phase lasting for the full simulation time, whereby the TDGs become significantly more compact and sustain significantly higher SFRs through compressive tides than the isolated model. 
None of the models is disrupted despite the unphysical extreme feedback, and none of the rotation curves achieves the high values observed in real TDGs, despite non-virial gas accretion phases.

\end{abstract}


\maketitle

\section{Introduction}\label{sec:intro}

Tidal dwarf galaxies (TDGs) form in the tidal debris of interacting galaxies. They are typically observed as gas-rich, star-forming objects that are still embedded in the tidal arm, that was pulled out from their progenitor galaxies. At this stage they are still young with estimated ages of less than 2 Gyr, as this is the time-scale in which tidal features are thought to disappear \citep{1995AJ....110..140H, 2010ApJ...717L.143M}. Young TDGs deviate from the luminosity-metallicity or mass-metallicity (MZ) relation \citep[e.~g.][]{1994AA...289...83D, 1998AA...333..813D, 2003AA...397..545W, 2009ApJ...705..723C, 2012AA...538A..61M}, as their gas originates from the disks of more massive galaxies. 

Fossil TDGs, that are not connected to a tidal structure anymore, are difficult to identify. At ages of several Gyr, they could lie closer to the MZ-relation, as their progenitors were more metal-poor and a larger fraction of the final metal enrichment was produced directly in the TDG. 

If a large fraction of TDGs would survive for several billion years, their contribution to the total dwarf galaxy (DG) population should be large. Various studies tried to estimate the percentage $f_{\mathrm{TDG}}$ of DGs that formed in a tidal scenario. Galaxy interaction simulations at high redshifts by \citet{2010ApJ...725..542H} and \citet{2012MNRAS.427.1769F}, where the interacting galaxies had gas fractions of 60 per cent and 80 per cent respectively, produced several long-living TDGs in each tidal arm. Following the assumption, that the TDG production rate was larger at high redshifts, when the progenitor galaxies had higher gas fractions, the number of fossil TDGs in the local Universe would explain the complete DG population and therefore $f_{\mathrm{TDG}}= 100$ per cent \citep{2000ApJ...543..149O}. \citet{2006AA...456..481B} presented a numerical parameter study of galaxy interactions and estimated $f_{\mathrm{TDG}} =  10$ per cent. Observational surveys of interacting galaxies in clusters led to an estimate of $f_{\mathrm{TDG}} =6$ per cent \citep{2012MNRAS.419...70K}, while \citet{2014ApJ...782...35S} derived $f_{\mathrm{TDG}} = 16$ per cent in groups and \citet{1996ApJ...462...50H} concluded that $f_{\mathrm{TDG}}$ could be as high as $50$ per cent in compact groups. All estimates are sensitive to the TDG production rate $P_{\mathrm{TDG}}$, the mean TDG lifetime $L_{\mathrm{TDG}}$, and the detection / resolution limit. These values will not only depend on the local environment but are expected to be redshift dependent. Detailed studies, both numerically and observationally, are necessary to constrain $P_{\mathrm{TDG}}(z)$ and $L_{\mathrm{TDG}}(z)$. This will lead at first to a more consistent determination of $f_{\mathrm{TDG}}$ and secondly to a prediction on fossil TDG properties, which helps to identify DGs with a tidal origin.

Interestingly, a most important characteristic of TDGs is that they cannot contain a large amount of dark matter (DM), 
if any, because the velocity dispersion of the DM halo of the interacting host's is too high to fit in the phase-space of TDGs and cannot 
be gravitationally captured in low-mass concentrations. This fact must stay in the focus to distinguish TDG survivors 
from CDM-formed DGs. With this respect it is worth noticing that \citet{2011AA...526A.114T} derived
M/L ratios of about 1-10 for Virgo cluster dEs in the mass range of $10^{8-9} \Msun$ and the brightness interval 
$M_I ~ -18 .. -19$. These values are simply explainable by pure stellar components.

Another striking signature of multiple TDG systems would be their correlated kinematical and 
structural properties, e.g. the concentration of their orbits to a thin plane as it exists for
the satellite galaxies around the MW and Andromeda. These so-called dwarf spheroidals (dSphs) form 
the faint end of the gas-free dwarf elliptical galaxies (dEs). Not only the coincidence of 95 per cent of the dSphs' 
orbital angular momentum vectors \citep{2012MNRAS.423.1109P}, but also their confinement to extremely thin planes, 
the MW ``disk of satellites" (DoS) \citep{2005AA...431..517K, 2009MNRAS.394.2223M,2010AA...523A..32K} and also the recently detected 
``Vast Plane of Satellites" around M31 by \citet{2013Natur.493...62I} favour their formation as correlated systems 
\citep{1995MNRAS.275..429L,2002ApJ...564..736P}. Outside the Local Group (LG) a similar configuration of dSphs was found by \citet{2013AJ....146..126C} in the M81 group. Very recently, an analysis by \cite{2014Natur.511..563I} showed that disks of satellites are not restricted to the Local Group but can be found around several galaxies in the local Universe. These structures are difficult to explain, if the DGs are the analogues of dark matter (DM) dominated DGs that are found in cosmological simulations \citep{2005AA...431..517K, 2007MNRAS.374.1125M, 2014MNRAS.442.2362P, 2014MNRAS.440..908P}, but the phase-space correlation arises naturally if they formed in one correlated tidal structure. Simulations of gas-rich galaxy interactions that were constrained to fit the observed stellar structures of Andromeda, produce several TDGs in their tidal arms that can explain the DoSs both around the MW and M31 \citep{2010ApJ...725..542H, 2012MNRAS.427.1769F, 2013MNRAS.431.3543H, 2014MNRAS.442.2419Y}.
Whether dSphs are DM-dominated as derived by several authors from the stellar kinematics is heavily debated \citep{1997NewA....2..139K, 2007NuPhS.173...15G, 2010MNRAS.406.1220W,2012MNRAS.424.1941C}.

Complementary to large-scale simulations that focus on the formation and therefore the production rate $P_{\mathrm{TDG}}$, we perform high resolution simulations of individual TDGs, to get additional insight into their lifetimes $L_{\mathrm{TDG}}$. In \citet{2014MNRAS.437.3980P}, hereafter: Paper I, we studied the response of TDGs to an initially high SFR for different stellar populations and found that TDGs can self-regulate their SF within 500 Myr without getting disrupted by the stellar feedback. \citet{2007AA...470L...5R} concluded the same with 2D simulations with a resolution down to 5 pc. In this work, we aim at further constraining the survivability of TDGs and therefore $L_{\mathrm{TDG}}$. As TDGs form and age in the vicinity of more massive galaxies, we focus here on the effect of the tidal field on the evolution of TDGs. Compressive tidal forces in interacting galaxies can enhance both the formation and the lifetime of stellar objects \citep{2008MNRAS.391L..98R, 2009ApJ...706...67R,2014MNRAS.442L..33R}.  
An observational proof that TDGs do not necessarily dissolve after a short time was found in a survey around early type galaxies \citep{2011MNRAS.417..863D}, which led to the discovery of the oldest, uniquely identified TDG with an age of around 4 Gyr by \citet{2014MNRAS.440.1458D}.

In this paper, we present the improved simulation setup, including new initial conditions and slightly modified stellar feedback routines (Sec.~\ref{sec:setup}), compared to the setup used in Paper I. In Sec.~\ref{sec:results} we present the simulation results and analyse the dynamical and chemical evolution both of the gaseous and the stellar component. The importance of the tidal field and other interesting findings from these simulations, such as the creation of a counter-rotating gaseous envelope, are summarised in Sec.~\ref{sec:conclusion}. In the appendix (Sec.~\ref{sec:appendix}), we present a short analysis of galaxy interaction simulations \citep[see][]{2010ApJ...725..542H,2012MNRAS.427.1769F, 2014MNRAS.442.2419Y} to guide and motivate the initial conditions used in this work.

\begin{figure}
\begin{center}
	\includegraphics*[width = \linewidth]{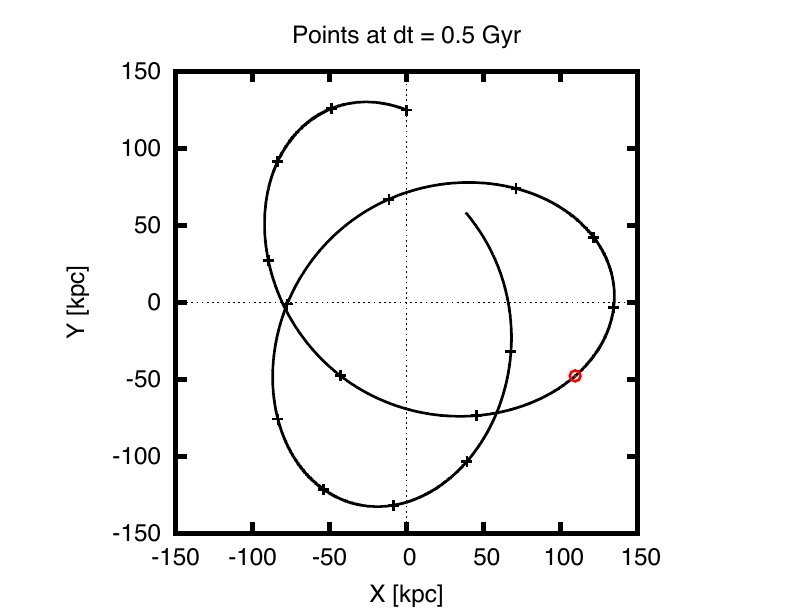}
	\caption{The orbit of the TDG simulations that include a tidal field is illustrated here. The coordinates are in the reference frame of the mass centre of the interacting galaxies. The position of the simulation box is indicated with crosses, starting at $(X,Y,Z) = (0,125,0)\,\mathrm{kpc}$, with one cross every 0.5 Gyr. The position of the TDG at the end of the simulation is indicated with a red circle.}
\label{fig:orbit}
\end{center}
\end{figure}

\section{Simulation setup} \label{sec:setup}

\subsection{Initial condition}

Kinematical studies reveal rotational signatures in young TDGs \citep{2007Sci...316.1166B, 2012MNRAS.427.2314L}. As detailed data on the 3D velocity structure are difficult to derive, we have analysed TDGs that formed in the galaxy interaction simulation presented in \citet{2012MNRAS.427.1769F}. The identified TDGs are all very young as the analysed snapshot shows the interaction 1.5 Gyr after the first encounter of the main galaxies. At this early stage, when the TDGs are still forming, half of the TDGs already have a rising and falling rotation curve. The other half shows a gradually rising rotation curve, which can mean that those TDGs have not yet collapsed into structures that are kinematically decoupled from the rest of the tidal arm (see Appendix). 

The simulations presented here start with a spherically symmetric, isothermal gas density distribution as in Paper I.  As a simplified approach that still resembles the properties of the TDGs found in observations and large-scale simulations, we set up TDGs with an initial rotation, which rapidly transforms the initial spherically symmetric gas distribution into a disk. As we were interested in the response to a high star formation rate (SFR) in Paper I, we started with cooler over-densities engulfed in a warm isothermal gas, which served as seeds of early star formation (SF). In this work, we focus on the long term evolution, so we start with a warm gas cloud of $1.2 \times 10^4\,\mathrm{K}$. Therefore, in this paper we start with a smooth, rotating gas cloud, while in Paper I the TDG already had an internal structure with cold, dense clumps initially. The central density is set to $n_H = 1\,\mathrm{cm}^{-3}$ resulting in a TDG with an initial radius of 6.4 kpc and an initial gas mass of $2.7 \times 10^{8}\,\Msun$. The box size is $(51.2\,\mathrm{kpc})^3$ with effectively $512^3$ grid cells, responding to a maximum resolution of $100\,\mathrm{pc}$. The simulation code is based on the Flash Code \citep{2000ApJS..131..273F}. The extensions that are necessary to run the TDG simulations were developed by the authors and presented in Paper I and this work.

In order to study the effects of the tidal field, the TDG moves within an external NFW \citep{1997ApJ...490..493N} potential $\Phi_{\mathrm{NFW}} (r)$ with 

\begin{equation}
	\Phi_{\mathrm{NFW}} (r) = - \frac{4 \pi G \rho_s r_{\mathrm{vir}}^3}{c^3 r} \, \mathrm{ln} \left(	1 + \frac{cr}{r_{\mathrm{vir}}}	 \right ) \, ,
\end{equation}

\noindent where $\rho_s$ is a characteristic density given by

\begin{equation}
	\rho_s = \frac{\rho_{cr} \Omega_m \delta_{th}}{3} \frac{c^3}{\mbox{ln} (1+c) - c/(1+c)} \, ,
\end{equation}

\noindent $\rho_{cr} = 3H^2 / (8 \pi G)$ the critical density of the universe, $\Omega_m$ the contribution of matter to the 
critical density and $\delta_{th}$ the critical over-density at virialization. 
As in Paper I, we use the virial radius $r_{vir} = 275\, \mbox{kpc}$ and the concentration parameter $c = 6.6$ for  $\Omega_m = 0.3$, $\delta_{th} = 340$ and $H_0 = 65\, \mbox{km}\,s^{-1}\,\mbox{Mpc}^{-1}$ from \citet{2008ApJ...684.1143X}. The virial mass of the DM halo is therefore defined by $M_{vir} = (4\pi /3) \rho_{cr} \Omega_{m} \delta_{th} r_{vir}^3 = 1.0 \times 10^{12}\, M_{\odot}$.

The orbit of the TDG is defined by the initial position and velocity within the reference frame of the interacting galaxies. At each time-step, the acceleration by $\Phi_{\mathrm{NFW}}$ determines the new position and velocity of the simulation box that includes the TDG, as well as the tidal field within the box. The TDG starts at $(X,Y,Z) = (0,125,0)\,\mathrm{kpc}$ with a velocity of $\vec{V} = (-100, 40, 0)\,\mathrm{km\,s}^{-1}$. With these initial values, the simulation starts, when the TDG is approaching the apo-centre at a distance of 135 kpc on a mildly eccentric (e = 0.36) orbit with a peri-centric distance of 64 kpc (see Fig.~\ref{fig:orbit}). On this orbit, the minimum velocity of the TDG at the apo-centre is $92.5 \,\mathrm{km\,s}^{-1}$ and the maximum velocity, at the peri-centre is $196 \,\mathrm{km\,s}^{-1}$. In the reference frame of the TDG, the accelerations caused by the tidal field are calculated for each time-step. Throughout the paper, coordinates in capital letters $(X,Y,Z)$ indicate the position relative to the mass centre of the interacting galaxies at $(X,Y,Z) = (0,0,0)$, while coordinates in lower case letters $(x,y,z)$ present the coordinates within the simulation box. Therefore the centre of the TDG at t = 0 is $(X,Y,Z) = (0,125,0)\,\mathrm{kpc}$ and $(x,y,z) = (0,0,0)\,\mathrm{kpc}$.

\begin{figure}
	\includegraphics*[width = \linewidth] {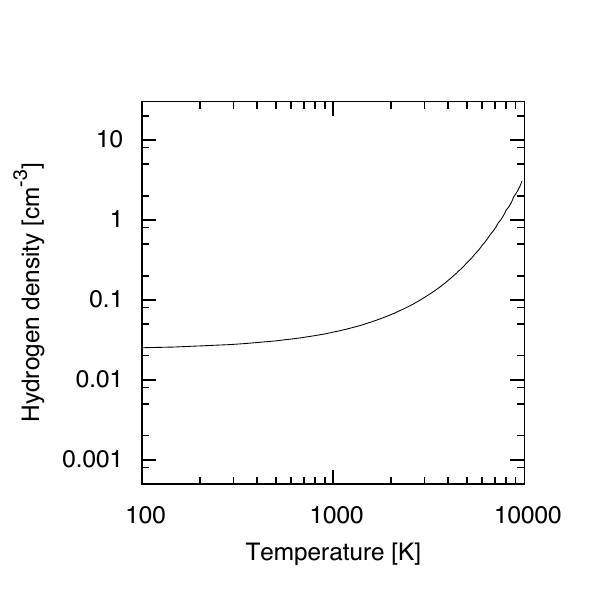}
	\caption{The SF threshold $\Psi(\rho, T)_{\mathrm{thres}}$ according to Eq.~\ref{eq:sbf_thres} with the used values for $c_{\mathrm{sf}}, \, M_{\mathrm{min}}, \, r_{\mathrm{gmc}}, \mathrm{\,and \,} \tau_{\mathrm{cl}}$ (see text). For higher temperatures the same $\Psi(\rho, T)_{\mathrm{thres}}$ leads to a higher density threshold than for lower temperatures.}
	\label{fig:SBF}
\end{figure}

\subsection{Star formation and stellar feedback}

Details on the SF and stellar feedback treatments can be found in Paper I. A few modifications were made that improve the accuracy of the included feedback processes. In this section we describe the slightly different SF threshold criteria (Sec.~\ref{sec:sfcriteria}), an advanced prescription to fully account for the feedback of winds from massive stars (Sec.~\ref{sec:windfeedback}), and an accurate calculation of the SNIa rate (Sec.~\ref{sec:snia}), which is especially important for long-term simulations. 

\subsubsection{Star formation criteria} \label{sec:sfcriteria}

As in Paper I, we use the stellar birth function $\Psi$ from \citet{1995AA...296...99K} to calculate the SFR in each grid cell over the time-step $\Delta t$, in dependence on gas temperature $T$ and density $\rho$:

\begin{equation}
	\Psi(\rho, T) = C_2 \rho^2  e^{-T/T_s} \, \mathrm{g\,cm^{-3}\,s^{-1}}
\end{equation}

\noindent
with $C_2 = 2.575 \times 10^8$ (in cgs units) and $T_s = 1000\,\mathrm {K}$ \citep{1995AA...296...99K}. In our simulations, a star particle can get mass from the surrounding gas within a sphere of radius $r_{\mathrm{gmc}}=150\,\mathrm{pc} = 1.5 \, \mathrm {dx}$, where dx is the minimum size of a grid cell.
Assuming a constant $\Psi(\rho, T)$ during the formation of a star cluster with a cluster formation time $\tau_{\mathrm{cl}}$, any threshold in $\Psi(\rho, T)$ can be related to a threshold in temperature and density as illustrated in Fig.~\ref{fig:SBF}. As in Paper I, we assume star formation to occur only in (embedded) clusters. 
We choose a resolution limit for individual star cluster of $M_{\mathrm{min}} = 100\,\Msun$ and a formation time of $\tau_{\mathrm{cl}} =  1\, \mathrm{Myr}$. As the feedback processes start only after the cluster has formed, a short cluster formation time ensures a rapid start of the feedback processes, which is necessary for the self-regulation of SF, especially for regions with high SFR densities. 

For all simulations presented here, the SF threshold is therefore:

\begin{equation} \label{eq:sbf_thres}
	\Psi(\rho, T)_{\mathrm{thres}} = c_{\mathrm{sf}} \frac{3 M_{\mathrm{min}} }{4 \pi r_{\mathrm{gmc}}^3 \tau_{\mathrm{cl}}}
\end{equation}

\noindent
where $c_{\mathrm{sf}} = 0.5$ is a constant to ensure that the minimum cluster mass is resolved even if $\Psi(\rho, T)$ varies while the cluster forms. In Fig.~\ref{fig:SBF} the relation between density and temperature for a given $\Psi(\rho, T)_{\mathrm{thres}}$ is shown.

Each stellar particle represents a star cluster with a mass that depends on the local conditions of the interstellar medium (ISM). After $\tau_{\mathrm{cl}} =  1\, \mathrm{Myr}$, a star particle is closed for further SF and starts with stellar feedback. As long as the SF criteria ($\Psi(\rho, T) > \Psi(\rho, T)_{\mathrm{thres}}$, converging flow) are fulfilled, new star particles can form, but further SF will be already influenced by the stellar feedback from previously formed star clusters. 
Star masses within one star cluster follow a \citet{2001MNRAS.322..231K} initial mass function (IMF). In Paper I, we have investigated different descriptions of the IMF, and their influence on the survivability and the metal enrichment of the TDG. Here, we focus on the maximum feedback case with filled IMFs (i.~e. an invariant IMF which is always sampled to the maximum stellar mass of $120 \,\Msun$ such that fractions of massive stars can occur) and very short cluster formation times for a quick feedback response. This can serve as an upper limit on the survivability of TDGs, as a more detailed treatment of the IMF, such as a truncation of the individual IMFs - in line with the IGIMF (integrated galactic IMF) theory - reduces the total feedback energy (see Paper I, Appendix).

\subsubsection{Stellar feedback}

The IMF is binned in 64 logarithmic mass bins. The lifetime of stars in each mass bin is dependent on the average star mass within the bin and the metallicity of the star cluster. Based on the tables from \citet{1998AA...334..505P}, the time at which the stars in a given mass bin end their stellar evolution is calculated. 

\paragraph*{Stellar wind feedback} \label{sec:windfeedback}

Lyman continuum radiation from massive stars ionises the surrounding ISM over the whole lifetime of the star within a sphere of radius $R_{\mathrm{S}}$, the Str\"omgren sphere. We have improved the wind feedback treatment used in Paper I to include a better sub-grid description of the individual Str\"omgren spheres.

In order to calculate the mass within a Str\"omgren sphere, the balance between the ionisation rate by the stellar Lyman continuum photons $L_{\mathrm{ly}}$ per star and the recombination rate for hydrogen in the surrounding ISM has to be calculated. For the resolution used in the simulation it can be assumed that the emitted photons are absorbed within the same grid cell (``on-the-spot-approximation"). Therefore we use a fit to the hydrogen case B recombination coefficient $\alpha_B$ \citep[][see their table 1]{1992ApJ...387...95F}: 

\begin{equation}
	\alpha_B = 2.6 \times 10^{-13} \left ( \frac{10^4\,\mathrm{K}}{T} \right )^{0.85}\, \mathrm{cm^3\,s^{-1}} \quad .
\end{equation}

\noindent
The Lyman continuum radiation emitted by massive stars is described by 

\begin{equation}
	L_{\mathrm{ly}} =  3.6 \times 10^{42} \left ( \frac{m_{\star} }{\Msun} \right )^4  \quad .
\end{equation}

\noindent
under the assumption that $L_{\mathrm{ly}} \propto m_{\star}^4$ (Hensler, priv. comm.) and that it matches the photon flux for high-mass stars \citep{2003ApJ...599.1333S}.

\noindent
During the lifetime of massive stars ($m_{\star} \ge 8\,\Msun$), the ISM with ambient electron density $n_e$ within a radius of 

\begin{equation}
	R_{\mathrm{S}} = \left ( \frac{3 L_{\mathrm{ly}} }{ 4 \pi \alpha_B n_e^2} \right )^{\frac{1}{3}}
\end{equation}

\noindent
is assumed to be fully ionised and set to a constant temperature of $T_{\mathrm{strom}} = 20\,000\, \mathrm{K}$. The total grid cell temperature is calculated as a mass-weighted average between the regular grid temperature and the temperature of the Str\"omgren spheres. While the remaining fraction of the cell mass can radiatively cool, the temperature of the fraction of the cell mass, that is within a Str\"omgren sphere stays constant until the star explodes as SNII.

\paragraph*{SNIa rate:}\label{sec:snia}

In Paper I, we used a constant binary fraction for all mass bins to calculate the number of SNIa events. Here, we simulate the evolution of TDGs for 3 Gyr, where an accurate SNIa rate is more important and therefore we improved the calculation of the binary fraction in our numerical treatment. 
We follow the description by \citet{2009AA...499..711R}, where the SNIa rate is given by:

\begin{align} \label{eq:snia_rate}
\begin{split}
	&R_{\mathrm{SNIa}}(t) = \\
	 &A \int_{m_{\mathrm{B,\,inf}}}^{m_{\mathrm{B,\,sup}}} \int_{\mu_{\mathrm{min}}}^{\mu_{\mathrm{max}}} f(\mu)  \, \psi(t - \tau_{m_2}) \, \xi_{\mathrm{IGIMF}} [m_{\mathrm{B}, \, \psi(t - \tau_{m_2})}] \, \mathrm{d} \mu \, \mathrm{d} m_{\mathrm{B}}  \, ,
\end{split}
\end{align}

\noindent
with the following variables and in brackets the values used in \citet{2009AA...499..711R}:

\begin{tabular}[h!]{lp{6cm}}
	$A$						&  normalisation constant (A = 0.09)\\ 
	$m_{\mathrm{B,\,inf}}$		&  minimum total binary mass ($m_{\mathrm{B,\,inf}} = \mathrm{max}(2 \cdot m_2(t), 3\,\Msun$) \\
	$m_{\mathrm{B,\,sup}}$		& maximum total binary mass ($m_{\mathrm{B,\,sup}} = 8\,\Msun + m_2(t)$) \\
	$\mu$					& ratio between the mass of the secondary star $m_2$ to the total binary mass $m_B$ ($\mu = \frac{m_2}{m_B} = \frac{m_2}{m_1+m_2}$) \\
	$\mu_{\mathrm{max}}$		& maximum ratio $\mu$ for equal masses for the primary star $m_1$ and the secondary star $m_2$ ($\mu_{\mathrm{max}}=0.5$) \\
	$\mu_{\mathrm{min}}$		& minimum ratio $\mu$ for the largest possible difference between the mass of the primary and the mass of the secondary star ($\mu_{\mathrm{min}} = \mathrm{max} \left [ \frac{m_2(t)}{m_B},\, \frac{m_B-8\,\Msun}{m_B} \right ] $) \\
	$f(\mu)$					& distribution function of mass ratios in binary systems ($f(\mu) \propto \mu^{\gamma}$, $\gamma = 2$) \\
	$\xi_{\mathrm{IGIMF}}$		& initial mass function; in case of the IGIMF description $\xi_{\mathrm{IGIMF}}$ is dependent on the SFR $\psi$ at the time when the star was born\\
	$\tau_{m_2}$ 				& lifetime of the secondary star with mass $m_2$ \\
\end{tabular}

\noindent
We use the distribution function for mass ratios in binary stars, $f(\mu) = 2 ^{1+\gamma}(1+\gamma)\mu^{\gamma}$ with $\gamma = 2$ \citep{1983AA...118..217G, 1986AA...154..279M, 2013MNRAS.435.2460B}. Notice that observations of binary systems seems to suggest smaller values of $\gamma$ \citep[see e.~g.][]{1991AA...248..485D}. However, we use this value of $\gamma$ because Galactic chemical evolution models adopting it, reproduce the ratios in the Milky Way rather well \citep{2009AA...501..531M}.

From the SNIa rate derived from Eq.~\ref{eq:snia_rate} the number of SNIa in each mass bin $N_{\mathrm{SNIa}}$ is derived. The total energy input is given by $E_{\mathrm {SNIa, tot}} = N_{\mathrm{SNIa}} \times \epsilon_{\mathrm{SNIa}} \times E_{\mathrm{SNIa}}$, where $E_{\mathrm{SNIa}} = 10^{51}\,\mathrm{erg}$ is the energy input per SNIa, and $\epsilon_{\mathrm{SNIa}} = 0.05$ is the fraction of SNIa energy that is depleted onto the ISM. The released material follows the SNIa yields from the W7 model of \citet{2004AA...425.1029T}, which is mapped back onto the grid.

\begin{figure*}
	\begin{center}
		\begin {minipage}[b]{0.24\linewidth}
		\includegraphics[width=\linewidth, bb = 117 80 387 350, clip]{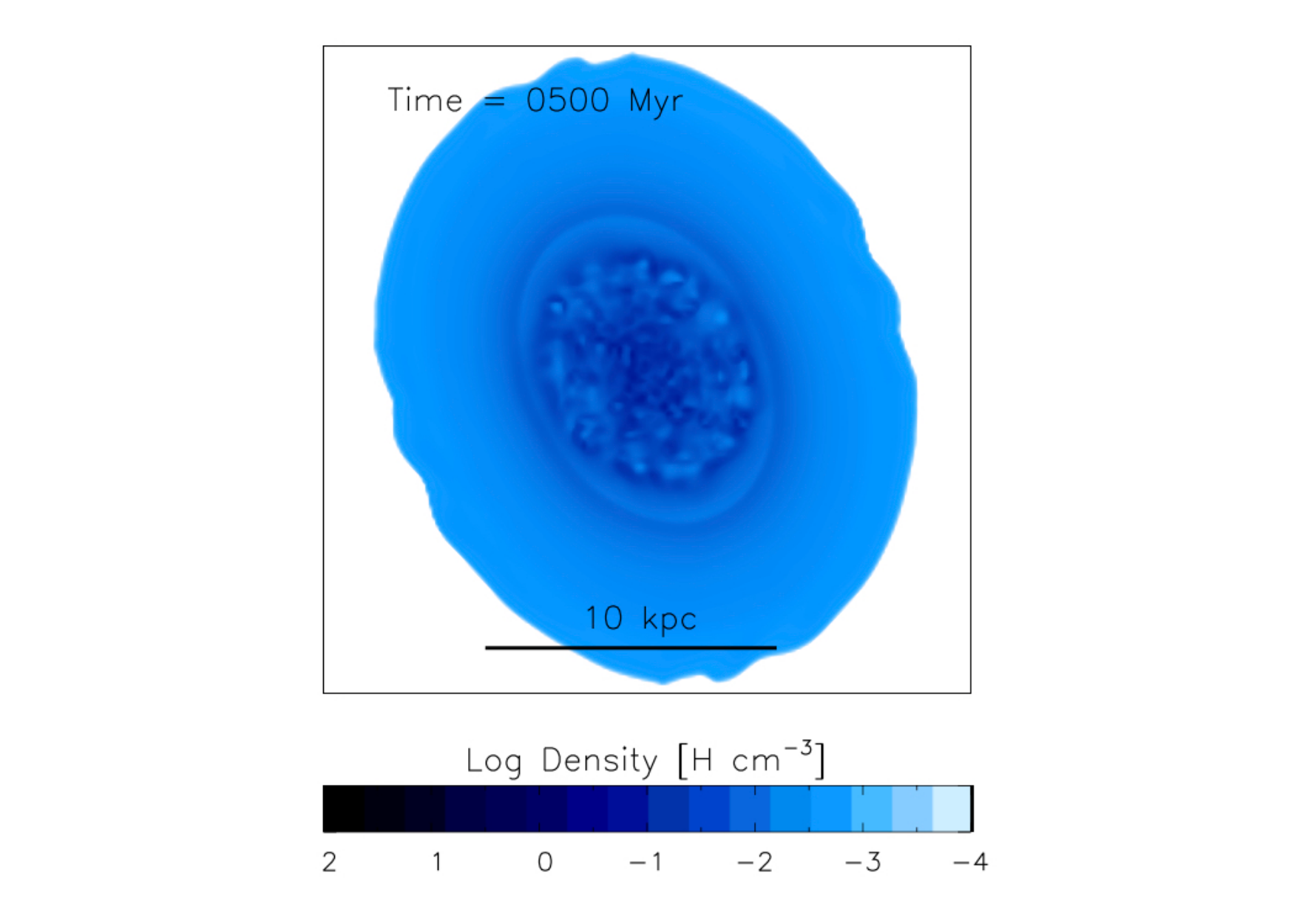}
		\end {minipage}
		\begin {minipage}[b]{0.24\linewidth}
		\includegraphics[width=\linewidth, bb = 117 80 387 350, clip]{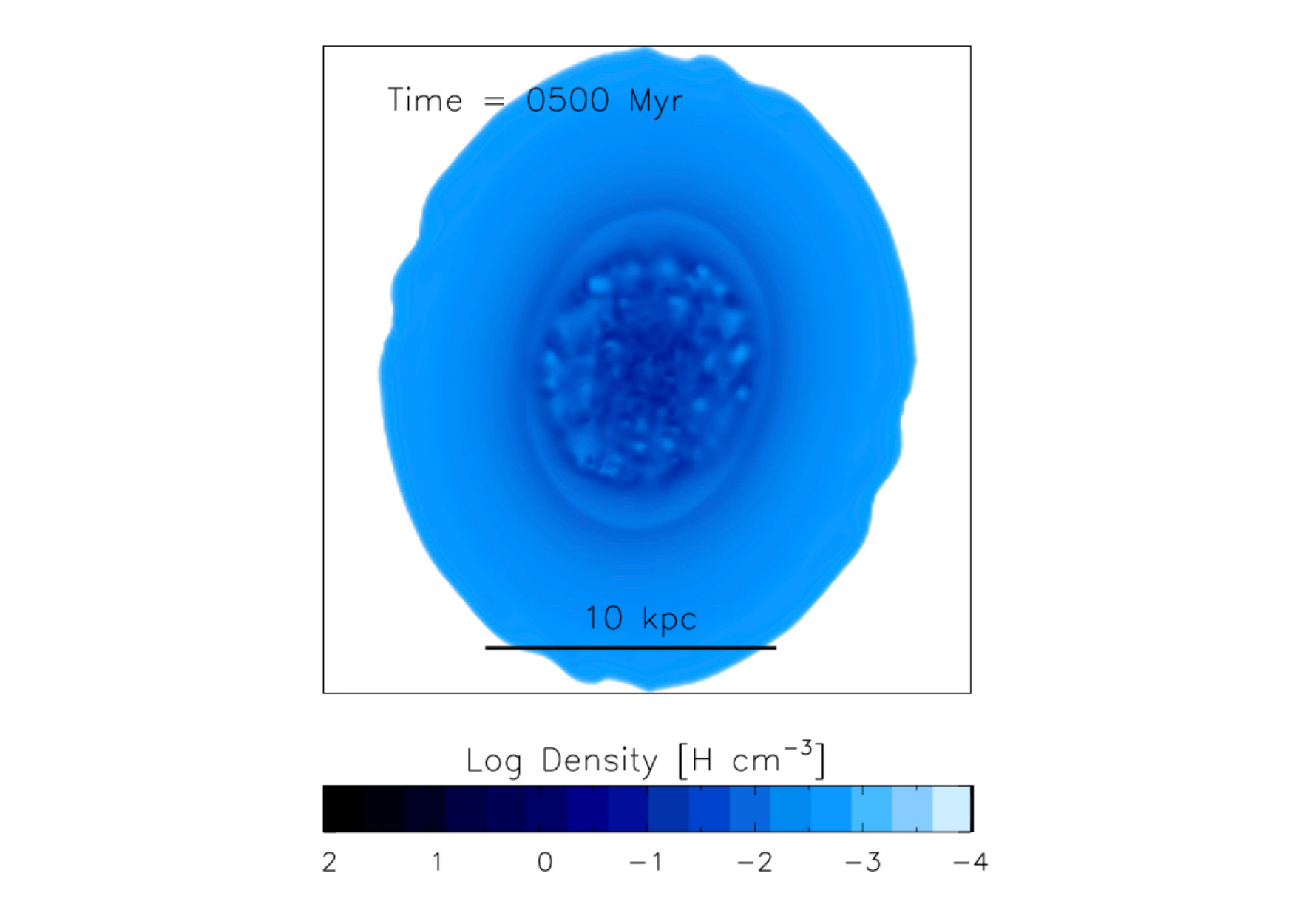}
		\end {minipage}
		\begin {minipage}[b]{0.24\linewidth}
		\includegraphics[width=\linewidth, bb = 117 80 387 350, clip]{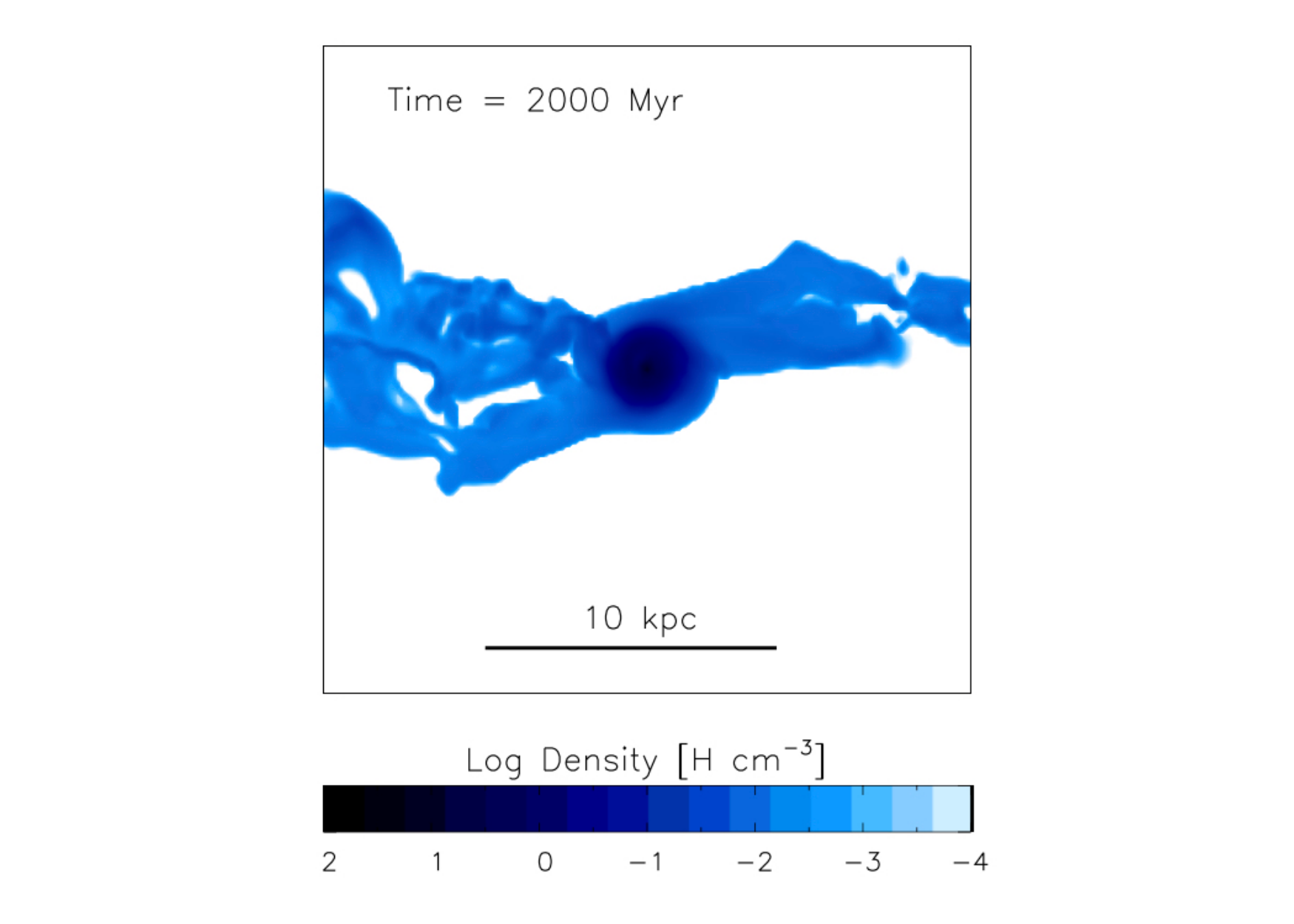}
		\end {minipage}
		\begin {minipage}[b]{0.24\linewidth}
		\includegraphics[width=\linewidth, bb = 117 80 387 350, clip]{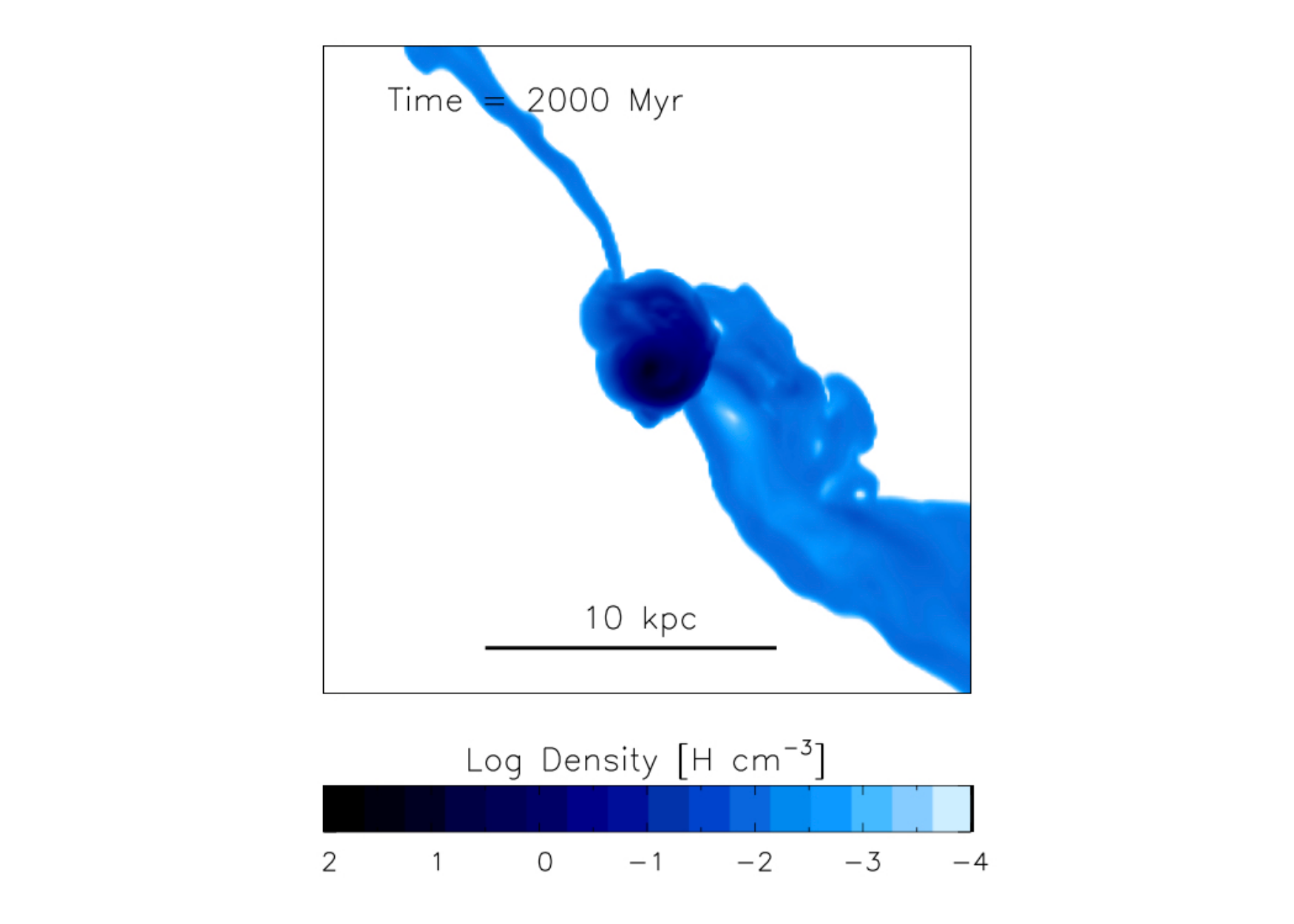}
		\end {minipage}		
		
		\begin {minipage}[b]{0.24\linewidth}
		\includegraphics[width=\linewidth, bb = 117 80 387 350, clip]{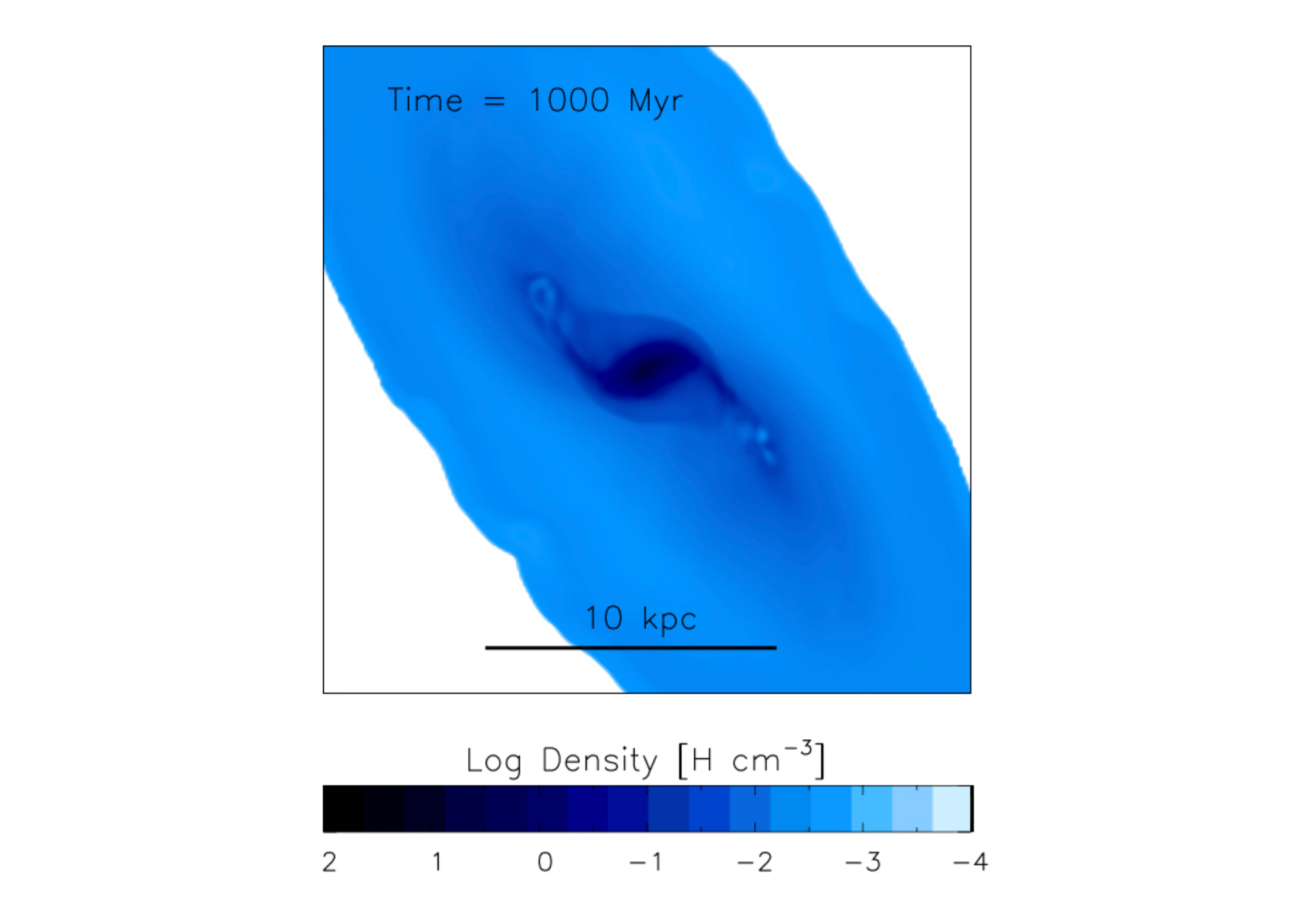}
		\end {minipage}
		\begin {minipage}[b]{0.24\linewidth}
		\includegraphics[width=\linewidth, bb = 117 80 387 350, clip]{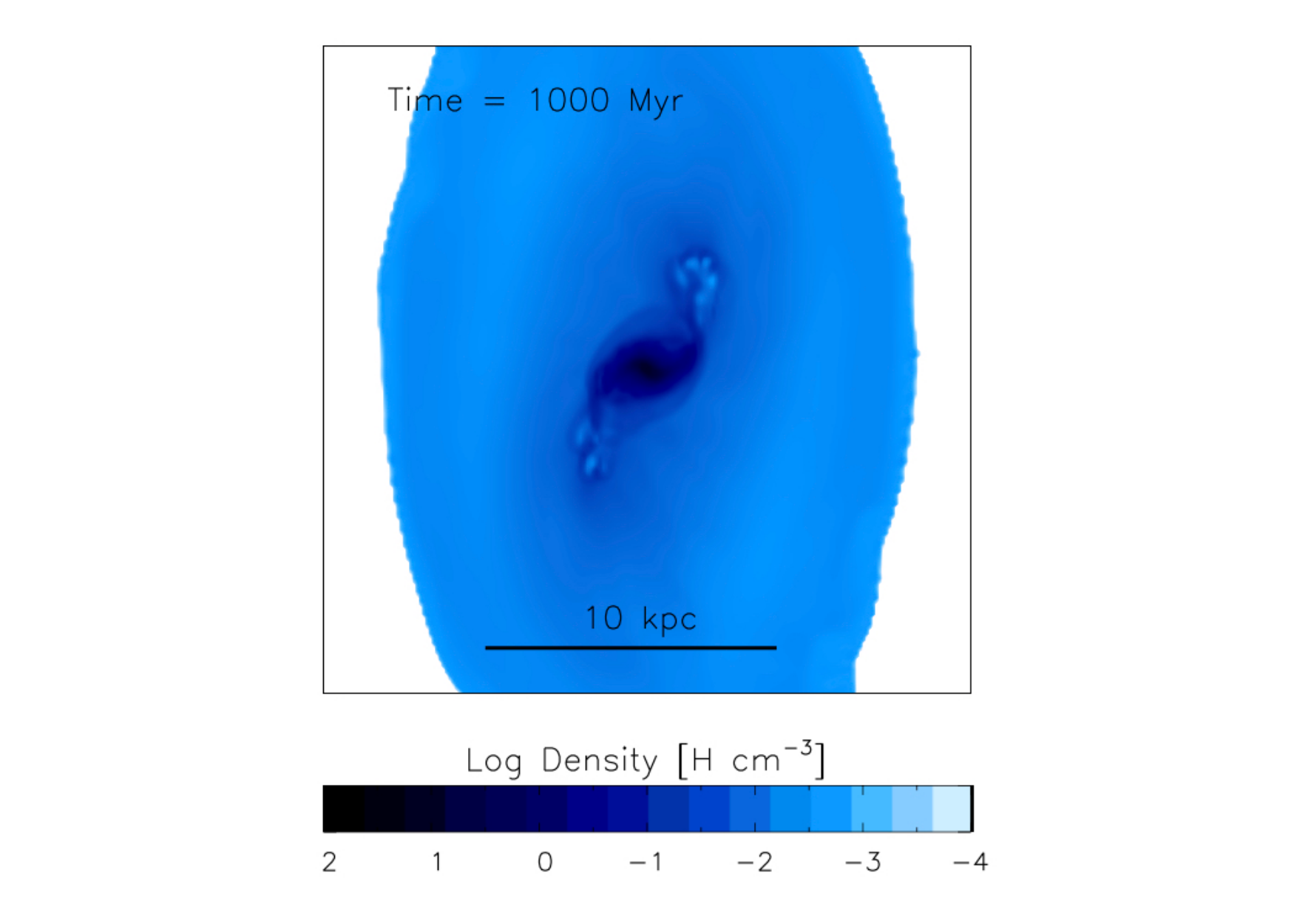}
		\end {minipage}
		\begin {minipage}[b]{0.24\linewidth}
		\includegraphics[width=\linewidth, bb = 117 80 387 350, clip]{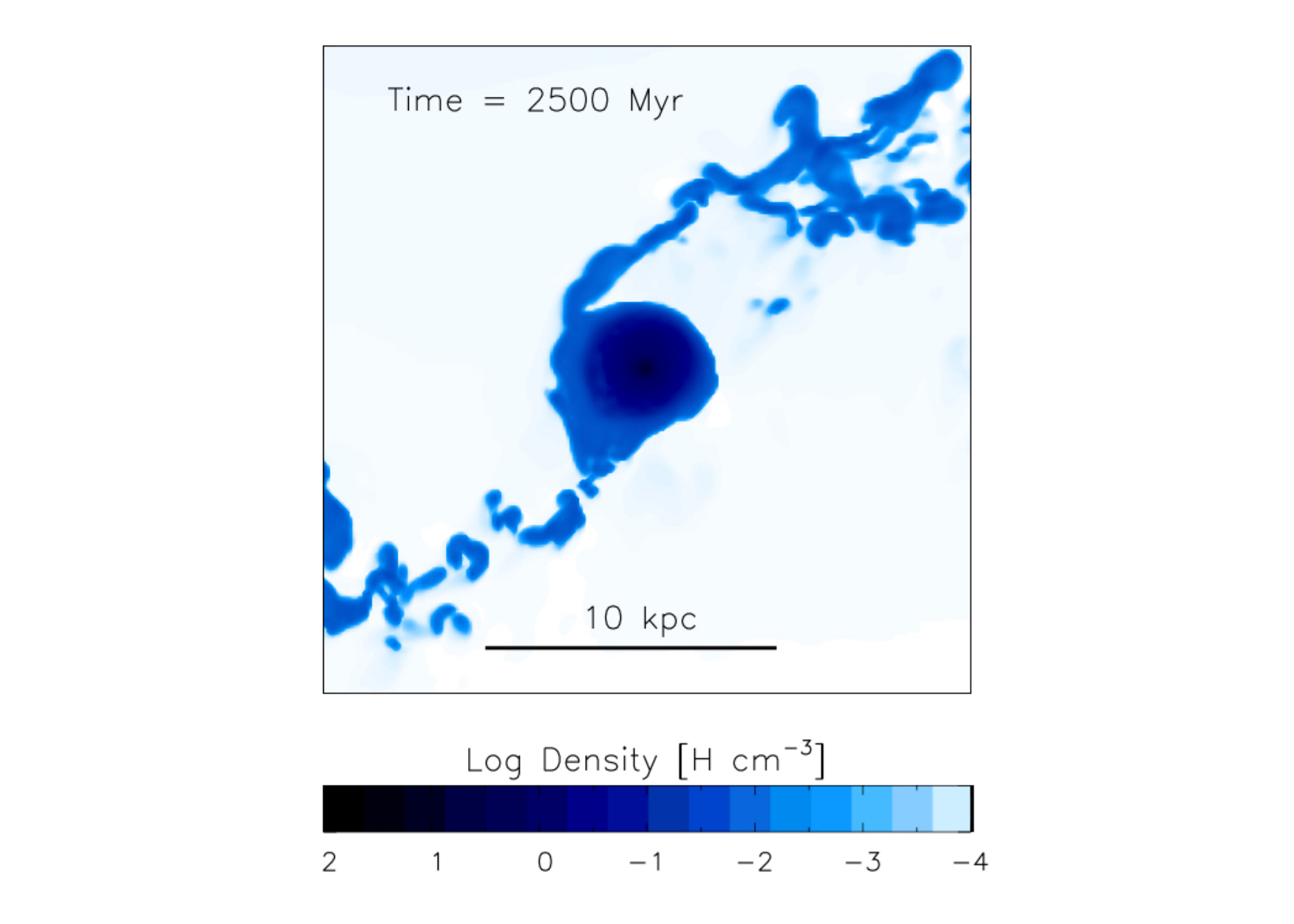}
		\end {minipage}
		\begin {minipage}[b]{0.24\linewidth}
		\includegraphics[width=\linewidth, bb = 117 80 387 350, clip]{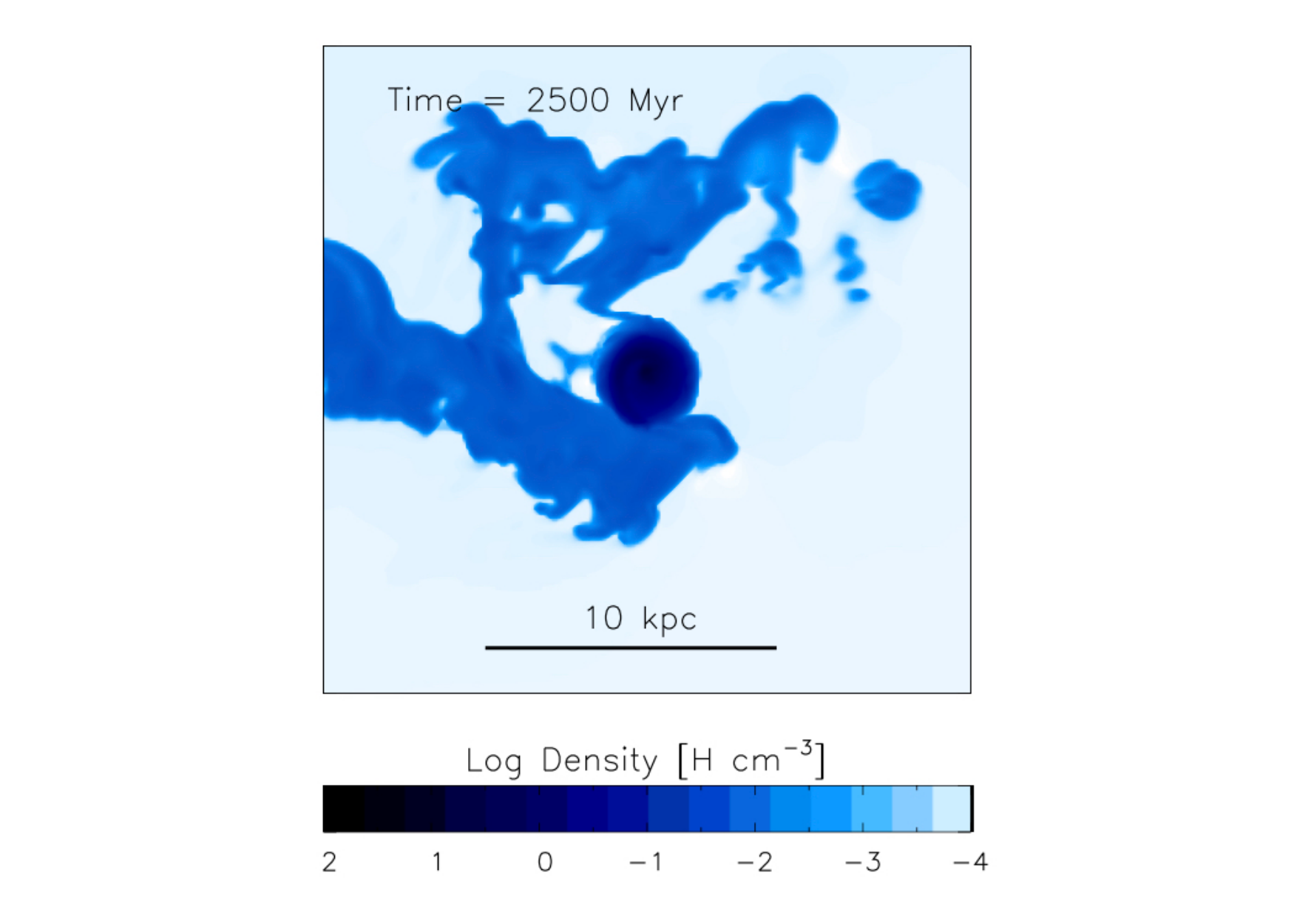}
		\end {minipage}		
		
		\begin {minipage}[b]{0.24\linewidth}
		\includegraphics[width=\linewidth, bb = 117 5 387 350, clip]{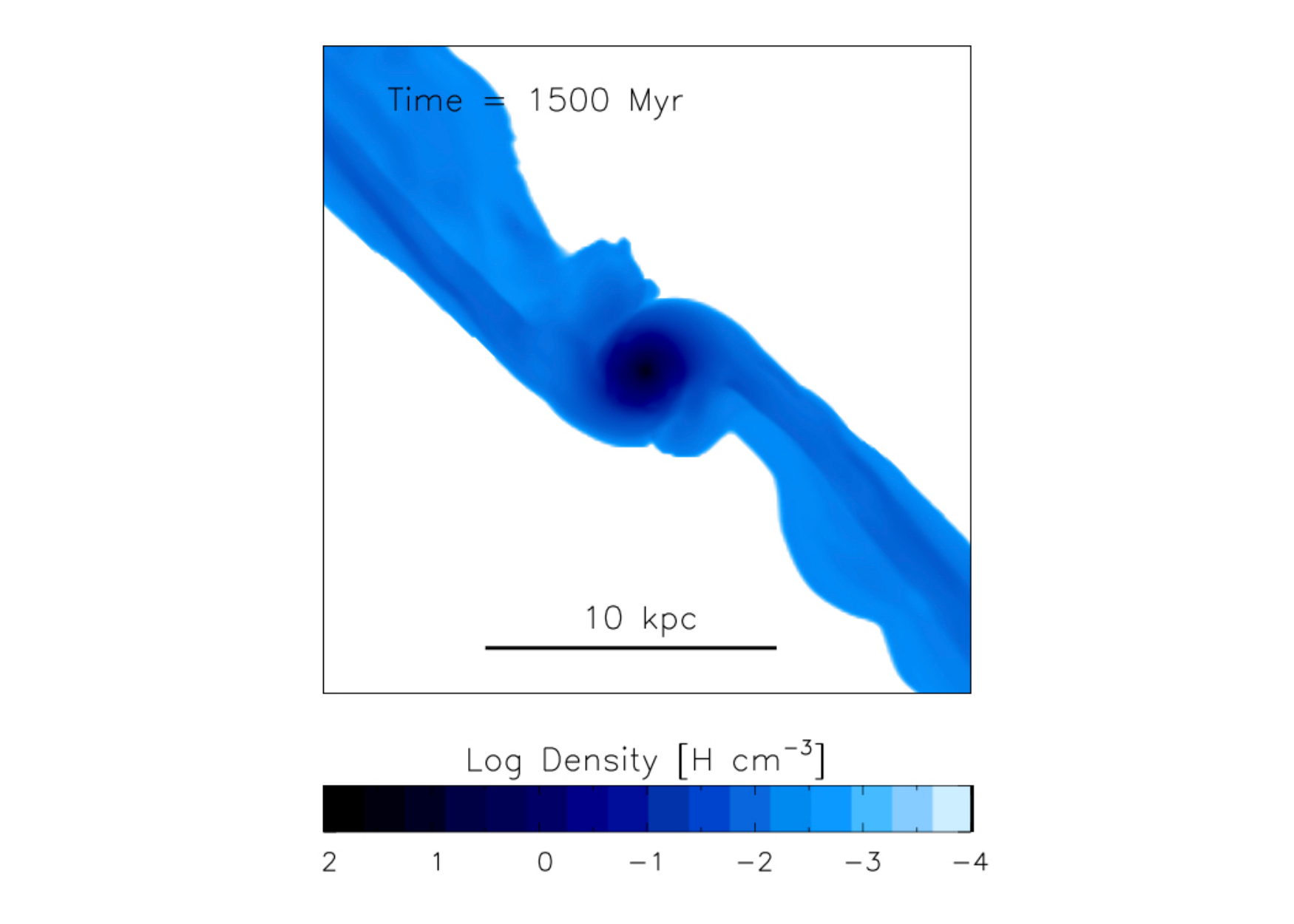}
		\end {minipage}
		\begin {minipage}[b]{0.24\linewidth}
		\includegraphics[width=\linewidth, bb = 117 5 387 350, clip]{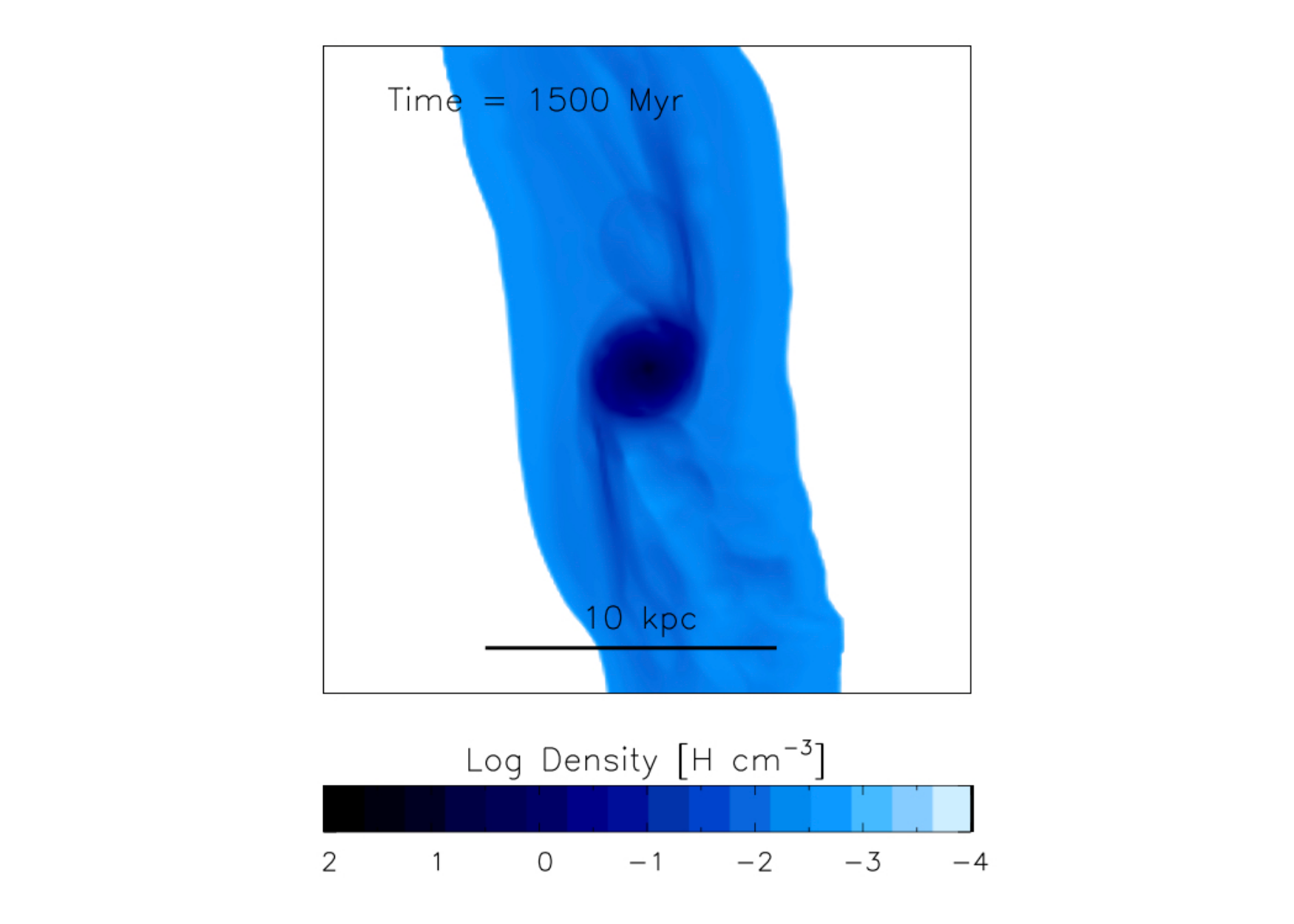}
		\end {minipage}
		\begin {minipage}[b]{0.24\linewidth}
		\includegraphics[width=\linewidth, bb = 117 5 387 350, clip]{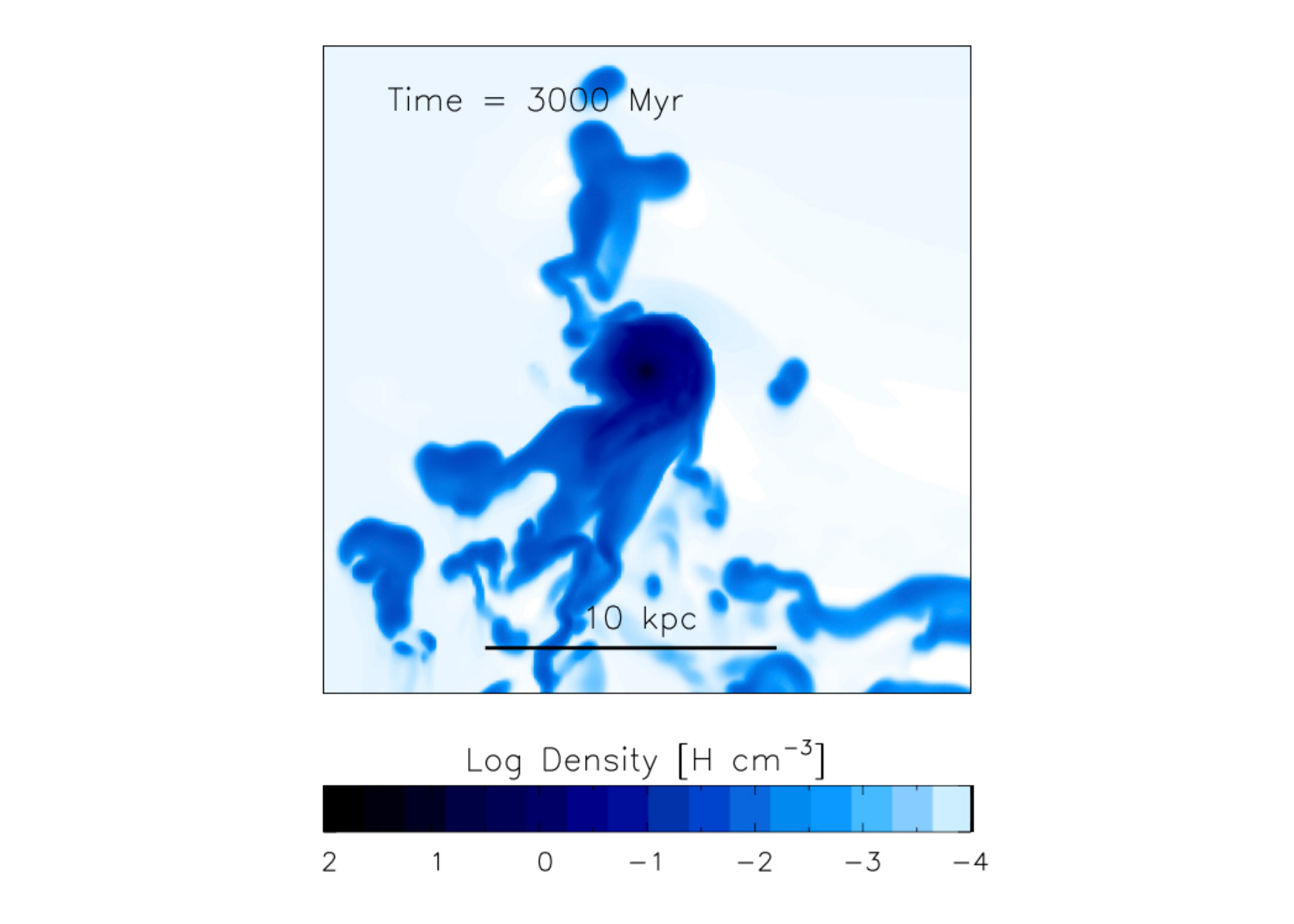}
		\end {minipage}
		\begin {minipage}[b]{0.24\linewidth}
		\includegraphics[width=\linewidth, bb = 117 5 387 350, clip]{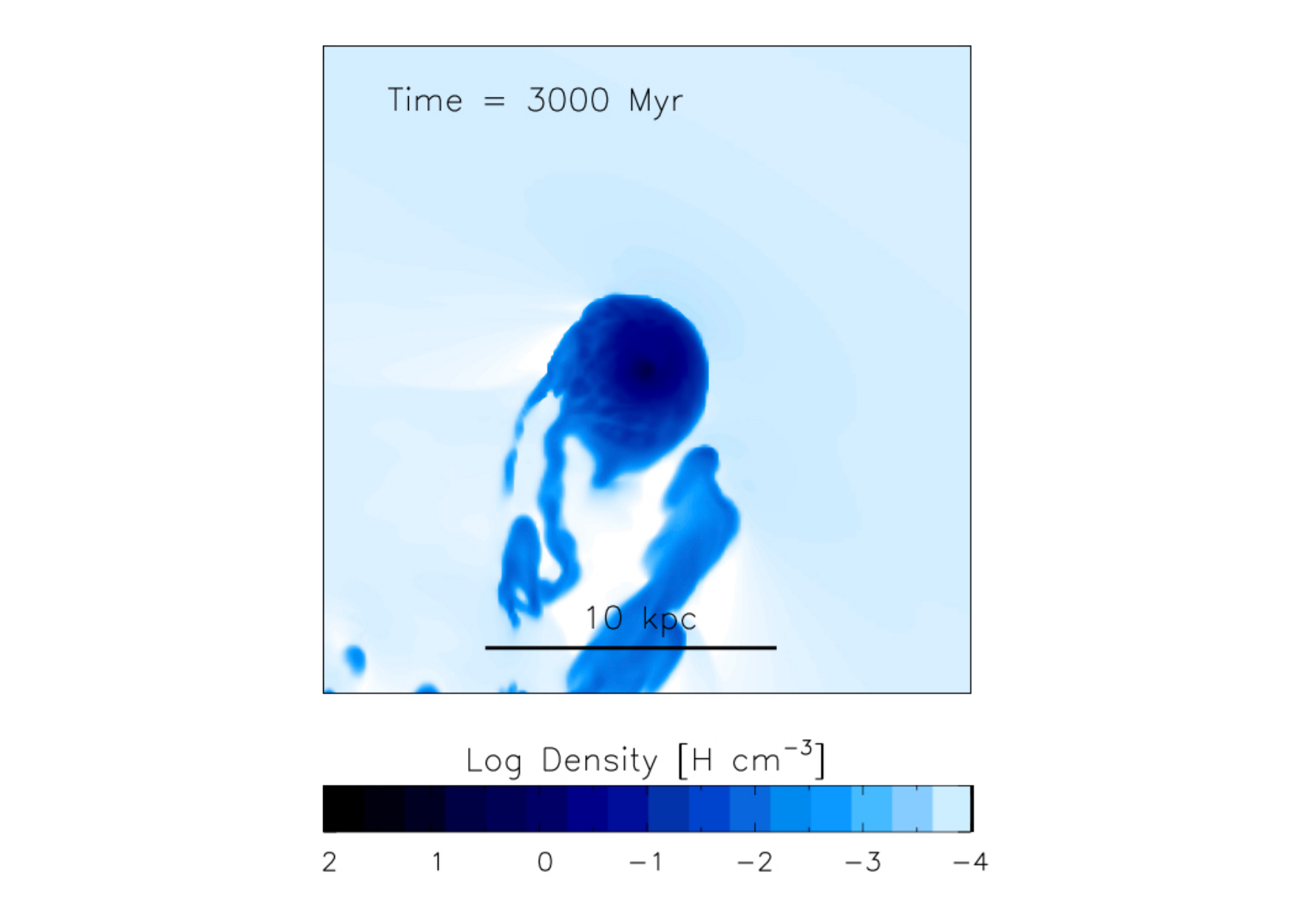}
		\end {minipage}	
		
		\caption{Gas distribution in the x,y plane and at z = 0 of TDG-p (column 1, 3), and TDG-r (column 2, 4) over time. Colour coded is the logarithmic hydrogen gas density.}
		\label{fig:gas}
	\end{center}
\end{figure*}    

\subsection{Boundary conditions} \label{sec:bc}

The simulation box of $(51.2\,\mathrm{kpc})^3$ follows the trajectory of the TDG. The ambient halo gas is initially set to a temperature of $10^6\,\mathrm{K}$ and a hydrogen number density of $2.2 \times 10^{-5} \,\mathrm{cm}^{-3}$. Only few X-ray observations of interacting galaxies exist \citep[e.~g.][around NGC 6240]{2013ApJ...765..141N}. Models of 
outflows that produce the hot gas halo by \citet{2006ApJ...643..692C} and hot gas shocks from halo collisions by \citet{2009MNRAS.397..190S} suggest that the densities at large distances are as low.  For the simulations presented here, we start with a TDG which is embedded in material that is co-moving with the TDG and therefore no initial relative velocity between TDG and the ambient medium. The accelerations within the simulation box are dominated by the tidal field and the self-gravity. Including an initial relative velocity would include also the effects of ram pressure stripping (RPS), as for example studied by \cite{2013MNRAS.436..839S}. 
Any relative streaming of this hot halo gas acting as ram pressure is however neglected here, because at such low gas densities with low relative velocities, stripping effects are of secondary order.

The boundary conditions (BCs) are set to the standard ``outflow"  boundary type in Flash. In this case, all four boundary cells get the same values from the adjacent cell within the computational domain (``zero-gradient" boundary conditions). This allows material as well as shocks to leave the computational domain but it does not distinguish between flows out of the domain and flows into the domain. In simulations, that include a tidal field, the TDG is compressed, especially when it is close to the peri-centre of its orbit. This compression leads to a net flow through the boundaries into the computational domain. As a result the density of the ambient medium increases during the simulation from initially $2.2 \times 10^{-5} \,\mathrm{cm}^{-3}$ to $1.3 \times 10^{-4}  \,\mathrm{cm}^{-3}$ at t = 3 Gyr and the temperature increases from $10^6\,\mathrm{K}$ to $2.7 \times 10^6\,\mathrm{K}$. 

It is possible to correct this effect and force the ambient medium to be constant over the whole simulation time, but as the TDG is embedded in more tidally expelled material which follows the same tidal field, a mass inflow onto the TDG at stages, where the tidal field is compressive, is more realistic. In the simulations presented here, the TDG can therefore accrete material from the surrounding tidal arm. For a more detailed investigation on the accretion rate from the tidal arm onto the TDG, the mass flow along the tidal arm has to be analysed from large-scale galaxy interaction simulations, as from \citet{2012MNRAS.427.1769F} and implemented as time-dependent BC type.

\section{Results} \label{sec:results}

Large-scale simulations of galaxy interactions show rotation patterns in their TDGs from a very early stage on. In the appendix we analyse the simulation by \citet{2012MNRAS.427.1769F} at 1.5 Gyr after the first encounter of the two progenitor galaxies. In their simulation, the disks of the progenitor galaxies are inclined, and the interaction process as well as the tidal arms are not confined to one plane. Therefore, the rotation of the TDGs within the tidal arm is also not restricted to the interaction plane. We calculate the angle $\alpha$ between the angular momentum vectors of the orbit and the internal spin of each TDG (see Appendix). For all TDGs, we found angles of less than $90\,\deg$, indicating a pro-grade rotation of the TDG relative to its orbit. 

As different galaxy interactions might lead to different TDG dynamics \citep[compare:] [on the formation of counter-orbiting tidal debris]{2011AA...532A.118P}, it is not yet clear, whether an initial pro-grade rotation of TDGs is a necessary condition from the tidal shearing field, or dependent on interaction parameter or the local velocity field during the clumping. Therefore, we started simulations with both an initially pro-grade ({\bf TDG-p}) and and initially retro-grade ({\bf TDG-r}) rotation. All other parameter are the same. As we want to investigate the role of the tidal field in the survival of TDG, we start a test TDG ({\bf TDG-t}) with the same initial conditions but without a tidal field. This allows to differentiate between effects that are artefacts from the chosen initial conditions or SF routines and real effects caused by the tidal field.

\begin{table}
\caption{Bound gas mass in $10^8\, \Msun$ for the simulation runs TDG-p and TDG-r.}
\begin{center}
\begin{tabular}{l|c|c}
\hline
\hline
t [Gyr]	& 	TDG-p	&	TDG-r 	\\
\hline
0.0		&	2.66		&	2.66		\\
0.5		&	2.58		&	2.60		\\
1.0		&	2.41		&	2.59		\\
1.5		&	1.82		&	2.34		\\
2.0		&	1.15		&	1.55		\\
2.5		&	0.99		&	0.98		\\
3.0		&	0.97		&	0.73		\\
\hline
\end{tabular}
\end{center}
\label{tab:boundgas}
\end{table}

\subsection{Gas distribution}

A time sequence of the initially spherically symmetric gas distribution of the simulation runs TDG-p and TDG-r is shown in Fig.~\ref{fig:gas}. 
In the beginning, at t = 500 Myr, all three simulations are still similar. The tidal field clearly stretches the TDG in the direction of the barycentre of the host galaxies and compresses it in the perpendicular direction.  As the orbit starts with an approach to the apo-center, where the tidal field is weakest, only a slight tilt and a slight elongation is noticeable. In the case of TDG-p the induced rotation by the tidal field has the same rotation direction as the initial internal rotation. Therefore the resulting angular momentum is a combination of both effects which becomes obvious in comparison with TDG-r, where the initial retro-grade rotation leads to different tilts of the system at the same times and positions within the orbit. The grainy structure in the central part of the TDGs at t = 500 Myr is caused by the stellar feedback. As the TDGs approach the peri-centre, the central part gets compressed while the material in the outskirts is lost. The analytical tidal radius $r_t$ 

\begin{equation}
	r_t = d \left [  \frac{M_{\mathrm{TDG}}}{M_{\mathrm{NFW}}(d) (3  + M_{\mathrm{TDG}} / M_{\mathrm{NFW}}(d) )}   \right ]^{1/3}
\end{equation}

\noindent
for the chosen NFW potential at a distance $d$ and the initial TDG mass of $M_{\mathrm{TDG}} = 2.66\times 10^8\,\Msun$ is 7.1 kpc at the apo-centre and 4.3 kpc at the peri-centre.
The evolution of the bound gas mass is presented in Table~\ref{tab:boundgas}. At each snapshot all gas cells where the sum of the internal energy and the kinetic energy is less than the potential energy are added to calculate the gaseous material that is gravitationally bound to the TDG.

\begin{figure}
\begin{center}
	\includegraphics[width=\linewidth]{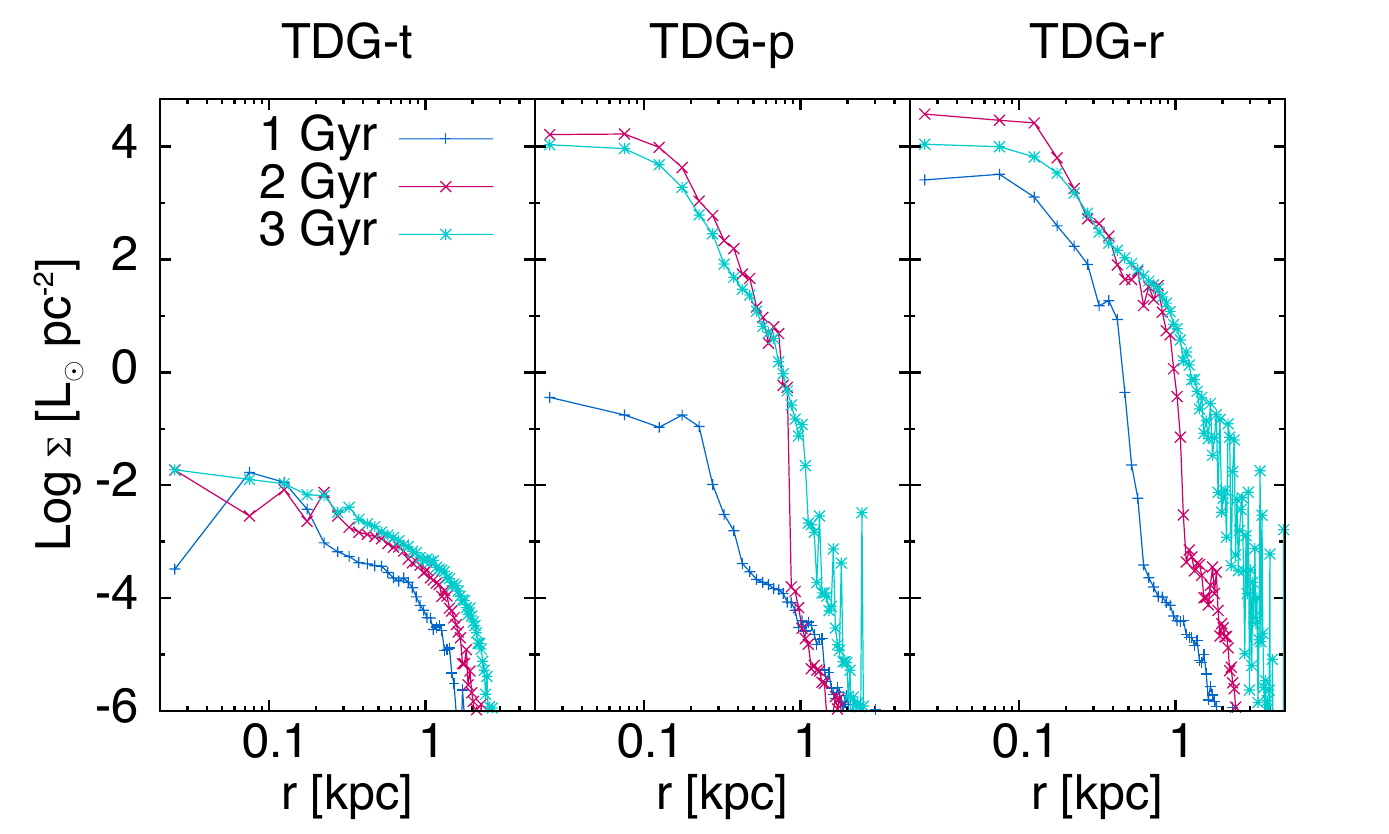}
\caption{Stellar component:  Face-on surface brightness profiles of TDG-t (left), TDG-p (middle), and TDG-r (right) for three snapshots at t =1, 2, and 3 Gyr. }
\label{fig:luminosity}
\end{center}
\end{figure}

\subsection{Stellar distribution}

 For every star particle we add the luminosities of all underlying mass bins that contain stars that have not yet reached the end of their stellar evolution following a mass-luminosity relation from \citet{2005essp.book.....S} of:

\begin{align}
\left ( \frac{L}{\mathrm{L}_{\odot}} \right ) \propto \begin{cases}
 (M_{\star}/\Msun)^{2.3}	,  	&  (M_{\star}/\Msun) < 0.43  \\
 (M_{\star}/\Msun)^{4} 	, 	&  0.43 \le (M_{\star}/\Msun) < 2 \\
 (M_{\star}/\Msun)^{3.5}	, 	& 2 \le (M_{\star}/\Msun) < 20 \\
 (M_{\star}/\Msun)		, 	& 20 \le (M_{\star}/\Msun)
\end{cases}
\end{align}

\noindent
The resulting total luminosities of each star cluster are binned in radial annuli with $\Delta r = 50\,\mathrm{pc}$ in the x-y plane and therefore a line-of-sight along z-direction (face-on) is assumed. The evolution of the surface brightness profiles $\Sigma(r) [\mathrm{L}_{\odot}\, \mathrm{pc}^{-2}]$ is shown in Fig.~\ref{fig:luminosity}. 

As seen in the gas distribution, the simulation runs that include a tidal field create compact stellar objects.  At the end of the simulation, the half light radius for TDG-t is approximately 550 pc while for both TDG-r and TDG-p the half light radius is close to 100 pc. The dynamics of the star particles are calculated by a particle-mesh technique implemented in Flash3.3. The masses of the particles contribute to the gravitational potential and the gravitational accelerations are mapped back from the grid onto the particles. This approach is very efficient for these already very CPU-expensive simulations, but it does not account for detailed stellar dynamics, such as the evolution of individual star clusters as well as close encounters between particles. The mentioned half-light radii therefore serve as a lower limit.
Interestingly, both TDGs in the tidal field show signatures of an additional central core component as found in the structure analysis of most dwarf elliptical galaxies in the Virgo cluster \citep{2014ApJ...786..105J}. \citet{2013MNRAS.429.1858D} compared the mass - radius relation of observed and simulated TDGs to early-type galaxies. In Fig. 2 of \citet{2013MNRAS.429.1858D}, the simulated TDGs presented in this work fall in the range of observed, young TDGs at a simulation time of t = 1 Gyr but at the end of the simulation, they are too compact or too mass-rich in stars compared to the observed TDGs. Follow-up studies with more sophisticated initial and environmental conditions will show if this is a necessary result of the tidal field for eccentric orbits.

\subsection{Star formation rate}

The SFR of the simulation runs TDG-t, TDG-p, and TDG-r are plotted in the top panel of  Fig.~\ref{fig:metallicity_new}. The simulation without a tidal field has an almost constant SFR over the full simulation length of 3 Gyr. A slight enhancement is visible around t = 1 Gyr, which is related to the chosen initial conditions. At this time, the gas cooling in the centre is very efficient which leads to a weak central collapse. This is quickly balanced by the stellar feedback of the newly formed stars and between t = 1.5 Gyr and t = 3 Gyr the SF has self-regulated to a constant SFR of approximately $2.5 \times 10^{-4}\,\Msun\,\mathrm{yr}^{-1} $. The star formation history (SFH) for the TDGs in the tidal field look very different. In both cases, the SFR is enhanced by 2.5 orders of magnitude after the apo-centre passage, with a moderate peak close to the peri-centre passage at 2.05 Gyr. Although the stellar feedback efficiently regulates the SF in the TDG without a tidal field, the compression of the tidal field is too strong and SF can continue for the remaining 2 Gyr with an exponential decay after the peri-centre. 
The stellar masses throughout the simulations are listed in Table~\ref{tab:stars}.

\citet{2014MNRAS.440.1458D} identified TDG around early-type galaxies and fitted a SFH to the spectral energy distribution of NGC 5551-E1. The best fit corresponds to an exponentially declining starburst 4 Gyr ago with SFRs of up to $0.08\,\Msun\,\mathrm{yr}^{-1}$. Especially TDG-p shows a very similar SFH.

\begin{table}
\caption{Stellar mass in $10^6\, \Msun$ throughout the simulation.}
\begin{center}
\begin{tabular}{l|c|c|c}
\hline
\hline
t [Gyr]	&	TDG-t	& 	TDG-p	&	TDG-r 	\\
\hline
0.0		&	0		&	0		&	0		\\
0.5		&	0.04		&	0.04		&	0.04		\\
1.0		&	0.10		&	0.18		&	1.37		\\
1.5		&	0.20		&	25.9		&	22.6		\\
2.0		&	0.29		&	66.3		&	77.8		\\
2.5		&	0.37		&	85.0		&	161.0	\\  
3.0		&	0.45		&	95.1		&	176.6	\\
\hline
\end{tabular}
\end{center}
\label{tab:stars}
\end{table}

\subsection{Rotation curves}

\begin{figure}
	\begin{center}
		\includegraphics[width=\linewidth]{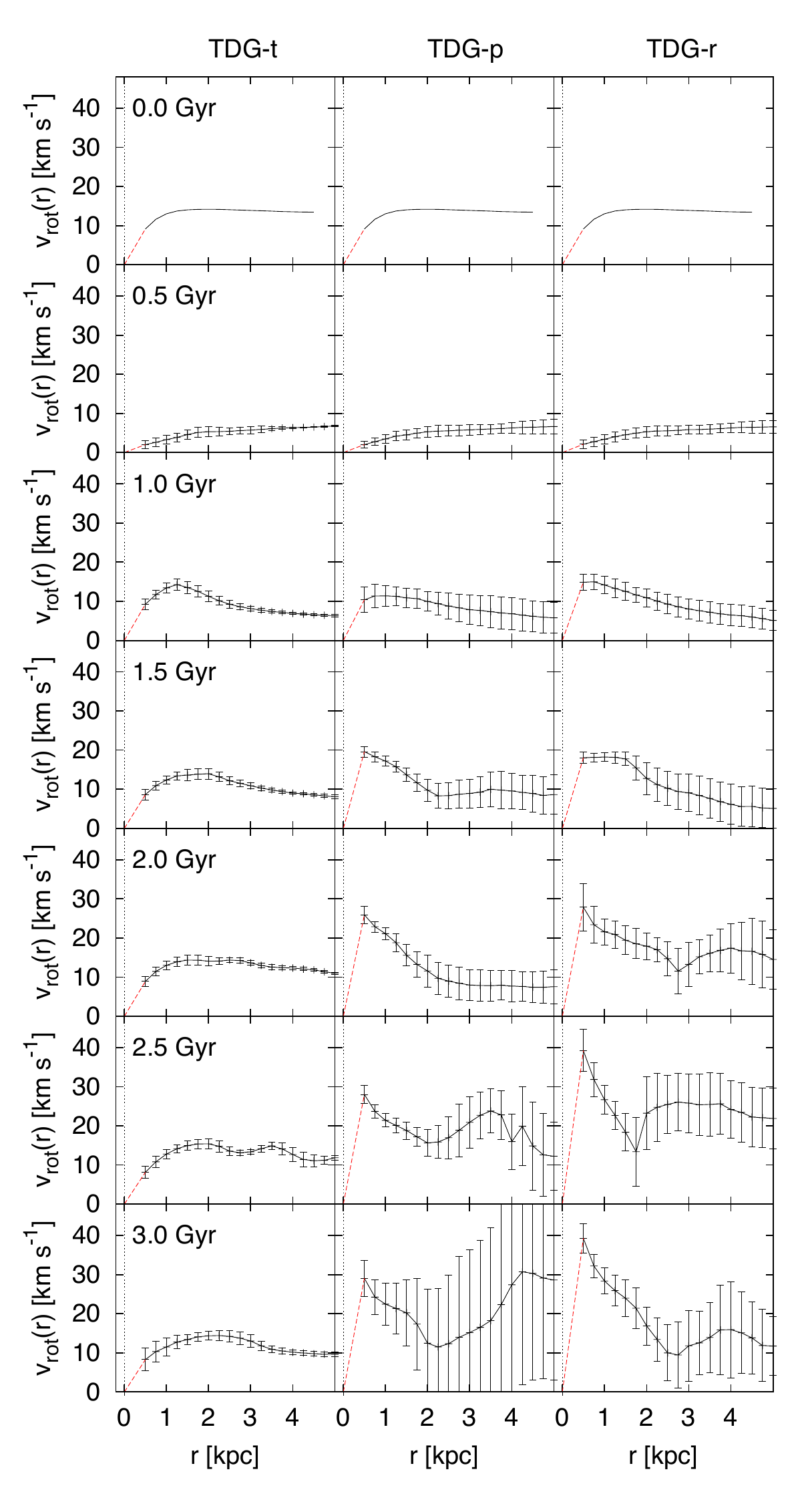}
		\caption{Rotation curves of the gaseous component of TDG-t (left panels), TDG-p (middle panels), and TDG-r (right panels). The absolute tangential velocities of all grid cells with densities $\rho > 10^{-26}\,\mathrm{g\,cm}^{-3}$ and within a vertical distance of $z = [-0.5, 0.5] \, \mathrm{kpc}$ to the x-y plane are averaged in radial distance bins of 0.5 kpc. The $1\sigma$ deviation within the bins is indicated with error-bars. The rotation curve for $r < 0.5\,\mathrm{kpc}$ (dashed line) is not resolved in this sampling.}
		\label{fig:rot_all}
	\end{center}
\end{figure}

\begin{figure}
	\begin{center}
		\rotatebox{90}{\includegraphics[height=0.48\linewidth,bb = 0 225 415 565, clip]{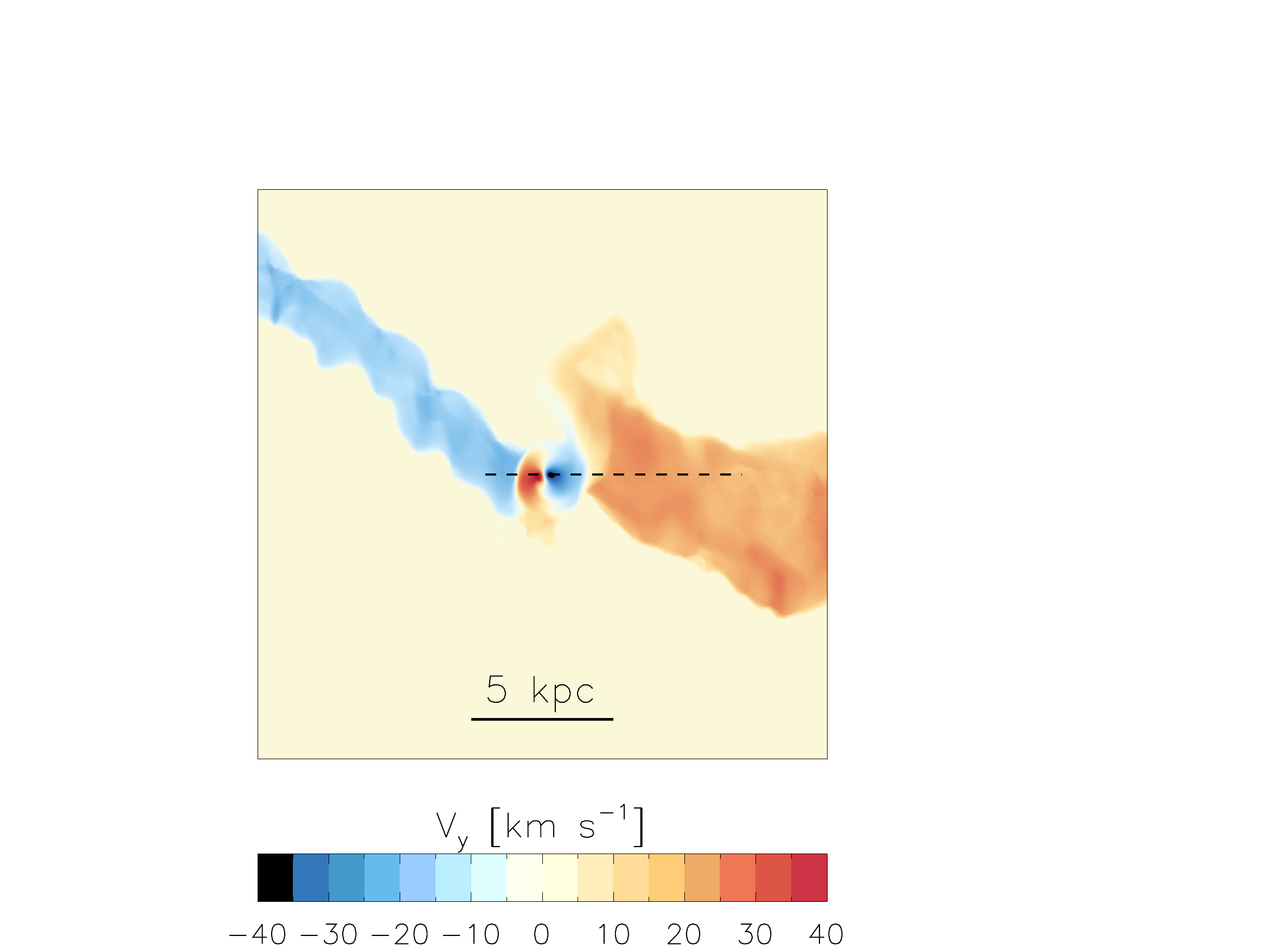}}
		\rotatebox{90}{\includegraphics[height=0.48\linewidth,bb = 0 225 415 565, clip]{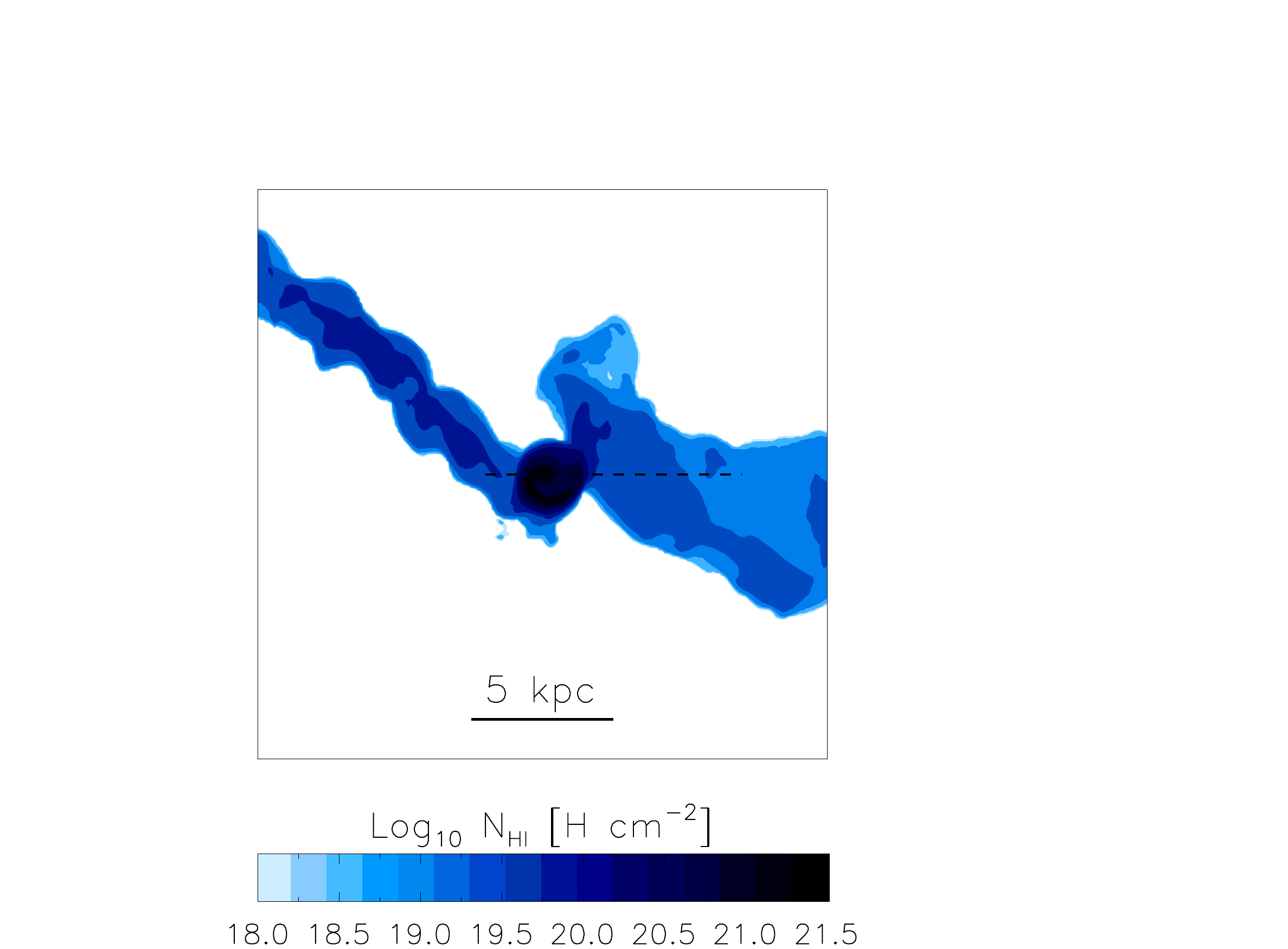}}
		\includegraphics[width=\linewidth]{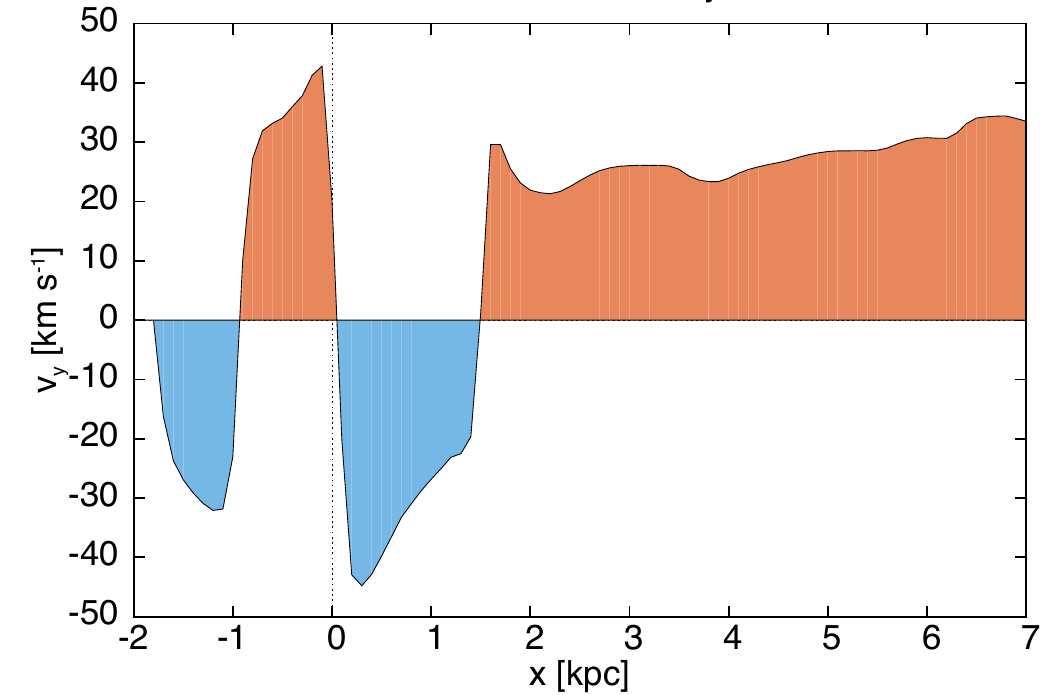}
		\caption{The top left panel shows the 2D velocity distribution of the gas in the XY plane through z = 0 for TDG-r at t = 2.2 Gyr. Colour coded is the velocity component in y direction ($v_y$), the direction of the ordinate in the top left figure. In the top right panel, the hydrogen column density for the same snapshot is plotted. The bottom figure shows $v_y$ along the dashed line indicated in the top figures. For illustration all velocities in grid cells with a density of $\rho < 10^{-26}\,\mathrm{g\,cm}^{-3}$ (ambient medium) are set to zero.}
		\label{fig:counter}
	\end{center}
\end{figure}

\citet{2007Sci...316.1166B} measured rotation curves of three TDGs in NGC 5291 and found flat rotation curves out to a radius of 5 kpc. The most common interpretation for flat rotation curves are in general either a DM component or a modification of the Newtonian law of gravity \citep{2007AA...472L..25G, 2010ApJ...718..380S}. In addition, an object could appear DM dominated if it is heavily perturbed or tidally heavily depopulated but an observer assumes dynamical equilibrium for the calculation of its dynamical mass \citep{1997NewA....2..139K}. As we perform long term simulations of TDGs, we can therefore study which effect the included tidal field has on the rotation curve of a TDG. As they cannot contain DM and therefore also the simulated TDGs are not embedded in their own DM sub-halo, and we also perform our simulations in standard Newtonian dynamics, since we can study if the tidal field and perhaps non-equilibrium gas accretion or gas flows near the forming TDGs, can lead to a flat rotation curve. 

The initial rotation velocity of each simulated TDG is set to 

\begin{equation}\label{eq:vel}
v(r) =  \sqrt{\frac{G\, M(r)} {r}} \; ,
\end{equation}

\noindent
where $r$ is the distance to the rotation axis of the TDG and $M(r)$ the enclosed mass, which leads to an axisymmetric rotation pattern. Along the orbit of the TDG within the tidal field of the interacting host galaxies, the TDG is stretched and compressed, which has an impact on the measured rotation curve. The rotation curves at $\Delta t = 0.5\,\mathrm{Gyr}$ intervals are shown in Fig.~\ref{fig:rot_all} for TDG-t (left panels), TDG-p (middle panels), TDG-r (right panels).

In the simulation without a tidal field (TDG-t), the initial gas sphere expands first radially, which is clearly noticeable in the rotation curve at t = 0.5 Gyr but keeps a stable rotation with a maximum rotation velocity close to the initial maximal rotation velocity of $14\,\mathrm{km\,s}^{-1}$ at all other times. Beside the rotation curve at t = 0.5 Gyr, all curves show declining rotation velocity at larger distances to the mass centre of the TDG. This is expected for a rotating, self-gravitating object in the standard Newtonian framework. 

Adding a tidal field has a large influence on the rotation of the TDG. After an initial expansion, the rotation curves of TDG-p (Fig.~\ref{fig:rot_all}, middle panels) and TDG-r (Fig.~\ref{fig:rot_all}, right panels) are significantly steeper than for TDG-t (Fig.~\ref{fig:rot_all}, left panels), as the gas clouds are compressed and the resulting TDGs are more compact than TDG-t. At the final snapshot, at t = 3 Gyr, TDG-t has a size of 13 kpc, while TDG-p has an extent of only 4 kpc. Even though the tidal field significantly perturbs the dynamics of the system, it does not flatten out the rotation curves in TDG-p and TDG-r. As in TDG-t, all snapshots after t = 0.5 Gyr show declining rotation curves. The rotation curves of these model TDGs encompass ages and masses and length scales which are similar to the three observed TDGs by \citet{2007Sci...316.1166B} \citep[see also][]{2007AA...472L..25G} such that the observed high and flat rotation curves of the real TDGs remain to be explained within the standard model of cosmology.

For the analysis of the rotation pattern of TDG-r, a radially binned rotation curve with the absolute tangential velocity does not allow to distinguish between the prograde rotation caused by the tidal field and the retro-grade rotation, which was set up in the initial conditions for this simulation run. The eccentric orbit around the interacting galaxies creates a torque on the TDG which is visible in the tilt of the tidal arms of the TDG in Fig.~\ref{fig:gas}. This torque leads to an angular momentum and therefore a rotation with the same orientation as the orbit. It is interesting to investigate if it is possible to flip the initially retro-grade rotation into a pro-grade rotation which for continuing SF could be observed as counterrotating stellar populations with well separated ages as observed in dwarf early-type galaxies in the Virgo cluster \citep{2014ApJ...783..120T}.

\begin{figure*}
	\begin{center}
		\includegraphics[width=\linewidth]{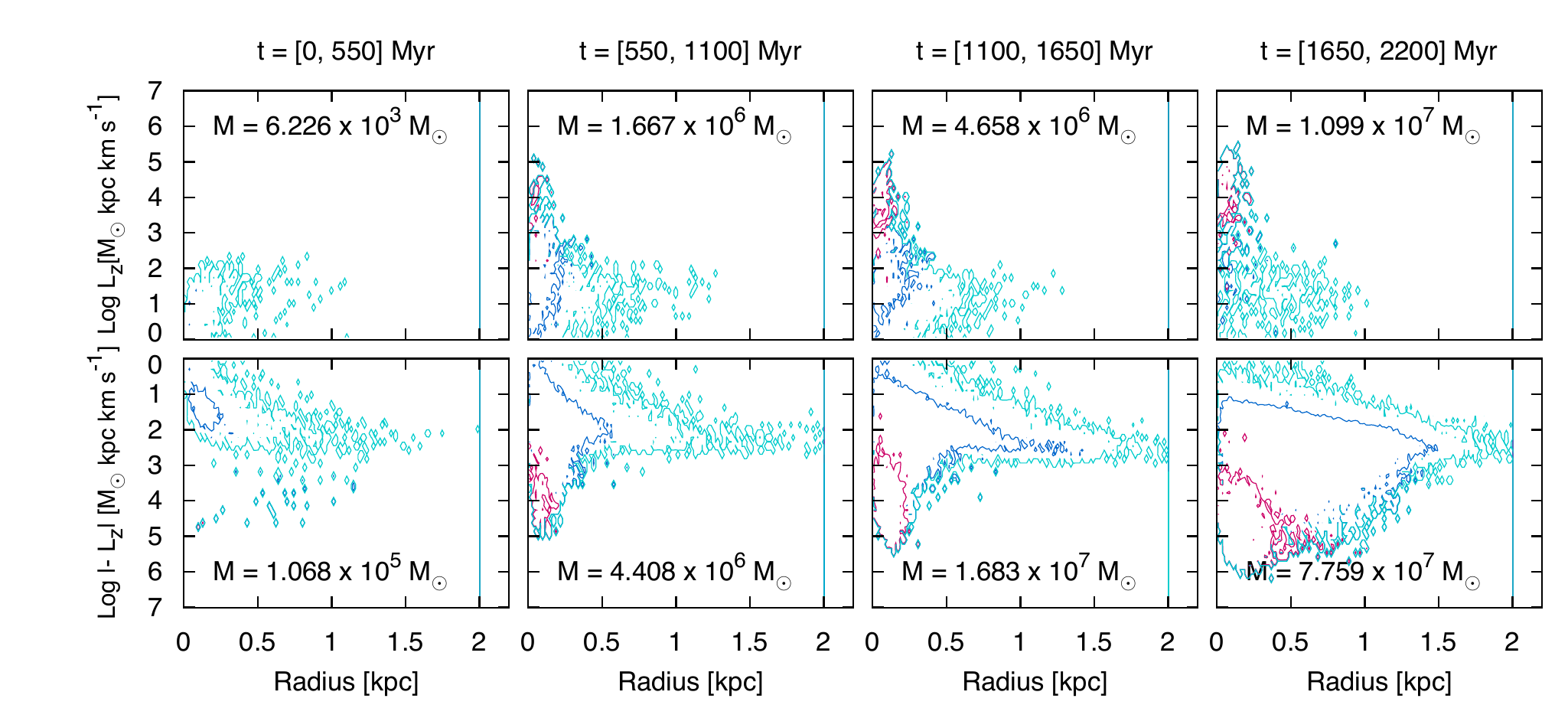}
		\caption{The figures show the rotation direction of the stellar component in the simulation TDG-r at t = 2.2 Gyr. For the velocity pattern of the gas phase at the same time, see Fig.~\ref{fig:counter}. In the top panels, only the stars with $L_z > 0$ (pro-grade) are indicated in dependence of their distance to the mass centre of the TDG, while in the bottom panels only stars with $L_z < 0$ (retro-grade) are displayed. In addition, the stars are separated in four sub-groups, depending on the time they were formed during the simulation. In the first column, stars that were formed in the first quarter of the simulation, t = (0, 550) Myr, are shown. The youngest stellar component at this snapshot, that formed between t = 1.65 Gyr and t = 2.2 Gyr, is presented in the fourth column. The contours indicate a relative mass density within this plot of 1, 100, and $10^4$. A region within the third contour therefore means that within one pixel of $\Delta r =15 \mathrm{pc}$ times $\Delta \mathrm{Log}_{10} (L_z ) = 0.12$, the total mass of stars that fall in this region is larger than $10^4\,\Msun$. The pixel size for the binning is arbitrary, but it illustrates the highly populated regions in the $r-L_z$ plot. }
		\label{fig:counterstars}
	\end{center}
\end{figure*}

Fig.~\ref{fig:counter} shows a snapshot of the velocity field of the gas phase in TDG-r at t = 2.2 Gyr right after the peri-centre passage. Clearly visible is the retro-grade rotating core with the typical declining rotation curve, while the outer gas parts, which were tidally stripped from the TDG show a pro-grade rotation caused by the torque through the tidal field. Fig.~\ref{fig:counterstars} shows the distribution of the rotation of the stellar component at the same snapshot. The x-axis is the radial distance of each star particle to the centre of the TDG. As both the initial rotation as well as the orbit of the TDG are in the X-Y plane we plot on the y-axis of Fig.~\ref{fig:counterstars} the z-component of the angular momentum vector. A point for each star particle in this plot would only create a large point cloud without much obvious information content. Therefore, we bin all stellar particles within this plot in pixels of $\Delta r =15 {\mathrm pc}$ and $\Delta \mathrm{Log}_{10} (L_z ) = 0.12$. For each of these pixels the stellar mass of star clusters in the $\Delta r ,\,  \Delta \mathrm{Log}_{10} (L_z )$ range are added and the contours show where most of the stellar mass is located in the r - $\mathrm{Log}_{10} (L_z )$ plot. The upper panels show the distribution of all particles with positive $L_z$, and therefore a pro-grade rotation while the lower panels show the distribution for star particles with negative $L_z$ and therefore retro-grade rotation. For the lower panels the logarithm of the absolute value of $L_z$ is shown. If the counter-rotation of the gas had already propagated to the stellar phase, a clear separation in $L_z$ in dependence of either the radius or the age of the stellar particles is expected. In Fig.~\ref{fig:counterstars} no distinct counter-rotation in the stellar phase is visible. While there is a population of stars in the central 0.5 kpc region with positive $L_z$ for all age ranges, no pro-grade rotating stars are found at $r > 1\,\mathrm{kpc}$, which would be expected if the in-falling prograde material has formed a significant population of stars. In this case, while the accreting low-density material doesn't form stars itself, it compresses the retro-grade rotating core which forms more retro-grade rotating stars. This can be seen in the rightmost bottom panel of Fig.~\ref{fig:counterstars}, where the youngest stars create a new over-density in the range $r \approx [0.5, 1.2]\,\mathrm{kpc}$ and $L_z = [10^3,10^5]\,\Msun\, \mathrm{kpc\,km\,s}^{-1}$. This region in the $r-L_z$ plot is not populated by older stars that were formed until $t = 1650\,\mathrm{Myr}$. 

In the simulation presented here, an initially retro-grade rotating gas cloud relative to its orbit in a tidal field leads to a counter-rotation in the gas phase. The stellar component that forms during the simulation is not clearly affected, as it is more concentrated in the centre, where the initial rotation of the gas still dominates.

\subsection{Metallicity}

Local stellar material feedback from AGB stars and supernovae (SNIa and SNII) enhances the element abundances of the surrounding ISM. In our setup stars with masses above $8\,\Msun$ enrich the ISM by SNII with delay times between 3 Myr ($120\,\Msun$ stars) and 45 Myr ($8\,\Msun$ stars). Stars with masses below $8\,\Msun$ release mass during the AGB phase or through SNIa events with delays between 45 Myr ($8\,\Msun$ star) and more than a Gyr (e.~g. 1.2 Gyr for a $2\,\Msun$ star). As the number of stars increases towards lower masses while the lifetime of stars increases, SNIa feedback gets increasingly important. 
Fe forms predominantly during a SNIa explosion while $\alpha$ elements, such as O, Mg, Si, and Ca form mostly during helium burning in the core of massive stars and are released during SNII events. Therefore the ratio [$\alpha$/Fe] is helpful to constrain the SFH in observations. 

\begin{figure}
	\begin{center}
		\includegraphics[width=\linewidth]{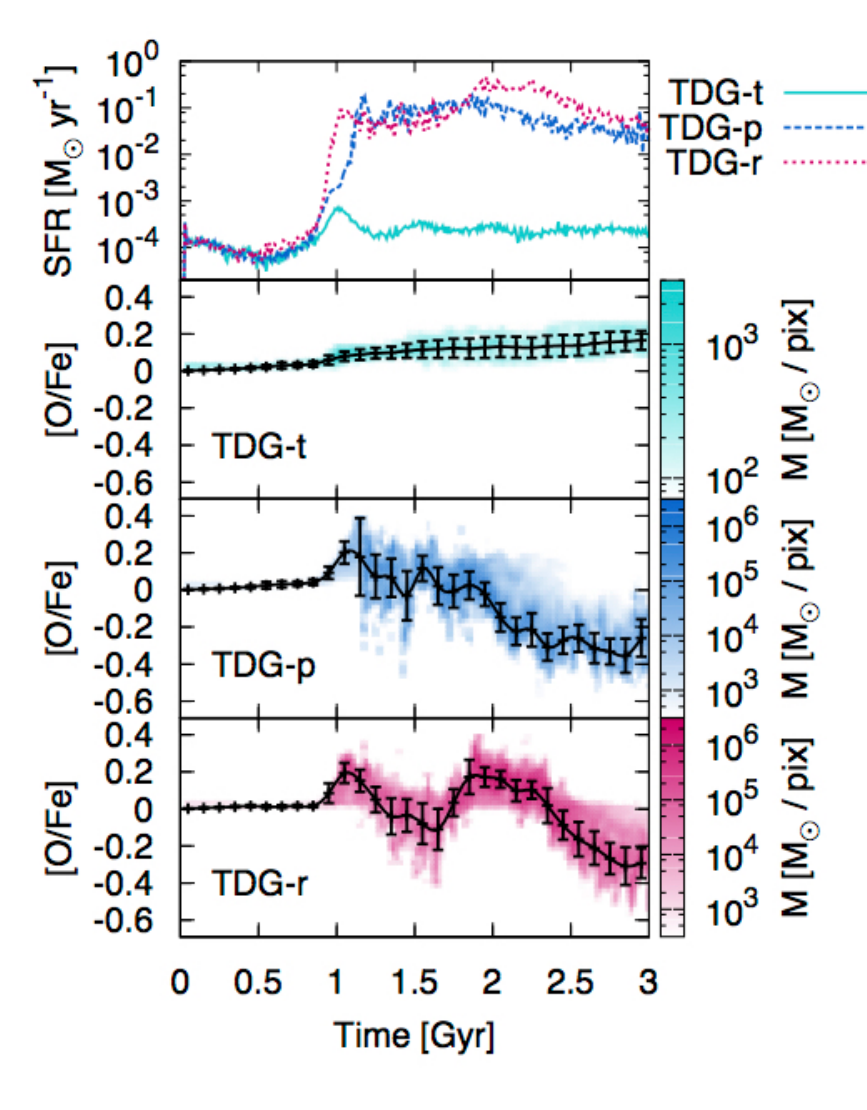}
		\caption{Top panel: SFRs for the simulation runs TDG-t (green, solid line), TDG-p (blue, dashed line), and TDG-r (red, dotted line). Bottom panels: Evolution of chemical abundances of the star clusters in TDG-t, TDG-p, and TDG-r (second, third, and fourth panel) as a 2D histogram of $[\mathrm{O/Fe}]$ of each star cluster against its formation time in the simulation. The stellar mass in each pixel of $\Delta t = 25\,\mathrm{Myr} \times \Delta [\mathrm{O/Fe}] = 0.025$ is colour-coded. The data points show the mass-weighted, logarithmic average and standard deviation error bars in time bins of 100 Myr. Note the different mass scale for TDG-t.}
		\label{fig:metallicity_new}
	\end{center}
\end{figure}

\begin{figure}
	\includegraphics[width=\linewidth]{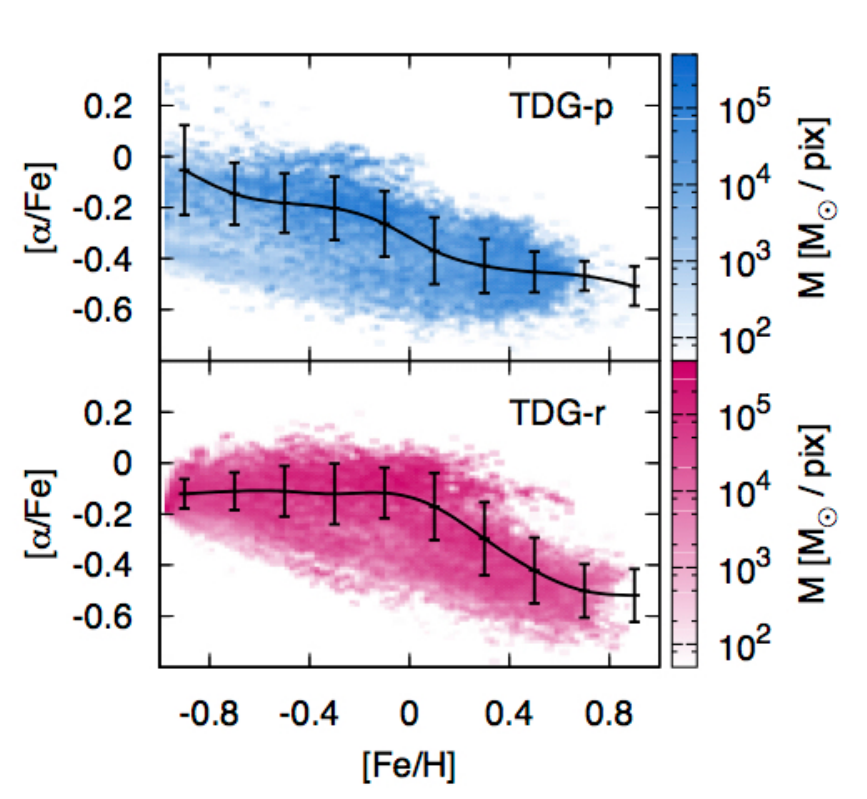}
	\caption{Alpha element (O + Mg + Si + Ca) abundance ratios for TDG-p (top panel) and \mbox{TDG-r} (bottom panel) of all star clusters at the end of the simulation. The colour-scale represents the stellar mass in each pixel of $\Delta [\mathrm{Fe/H}] = 0.02 \times \Delta [\alpha \mathrm{/Fe}] = 0.012$. The data points show the mass-weighted, logarithmic average and standard deviation in bins of 0.2.}
	\label{fig:alphaFe}
\end{figure}

The simulation starts with a pure gas cloud with $0.1\,\mathrm{Z}_{\odot}$ and solar abundance ratios, and therefore [O/Fe]~=~0 and [Fe/H]~=~-1.
In Fig.~\ref{fig:metallicity_new} the SFHs for the simulation runs TDG-t, TDG-p, and TDG-r are shown above the resulting [O/Fe] for each run. Clearly visible is the response of the stellar abundance ratios in TDG-p to the SF peak at 1 Gyr. While the [O/Fe] increases first (SNII), it decreases below the initial value after t = 2 Gyr, caused by Fe release of the delayed SNIa events. TDG-r shows a similar behaviour but with an additional [O/Fe] increase around 2 Gyr, related to a second SF peak at that time. After 3 Gyr, TDG-p and TDG-r both have abundance ratios of [O/Fe] $\approx -0.3$. In Fig.~\ref{fig:alphaFe}, the relation between [$\alpha$/Fe] and [Fe/H] for all star clusters at the end of the simulation is plotted. The slope as well as the ``knee" seen in TDG-r at [Fe/H] are reminiscent of observations of DGs \citep[compare e.~g.][]{2011ApJ...727...78K}, although the pre-enrichment leads to higher values of [Fe/H]. 

In general, galaxies follow a relation between their stellar mass and their metallicity (MZ-relation), where more massive galaxies are more metal-rich. 
Recently, \citet{2013ApJ...779..102K} fit the MZ-relation for DGs in the LG with

\begin{equation}  \label{eq:kirby}
	[\mathrm{Fe/H}]_{\mathrm {obs}} = (-1.69 \pm 0.04) + (0.30 \pm 0.02) \log_{10} \left( \frac{M_{\star}}{10^6\,\Msun}  \right)
\end{equation}

\begin{figure}
	\begin{center}
		\includegraphics[width=\linewidth]{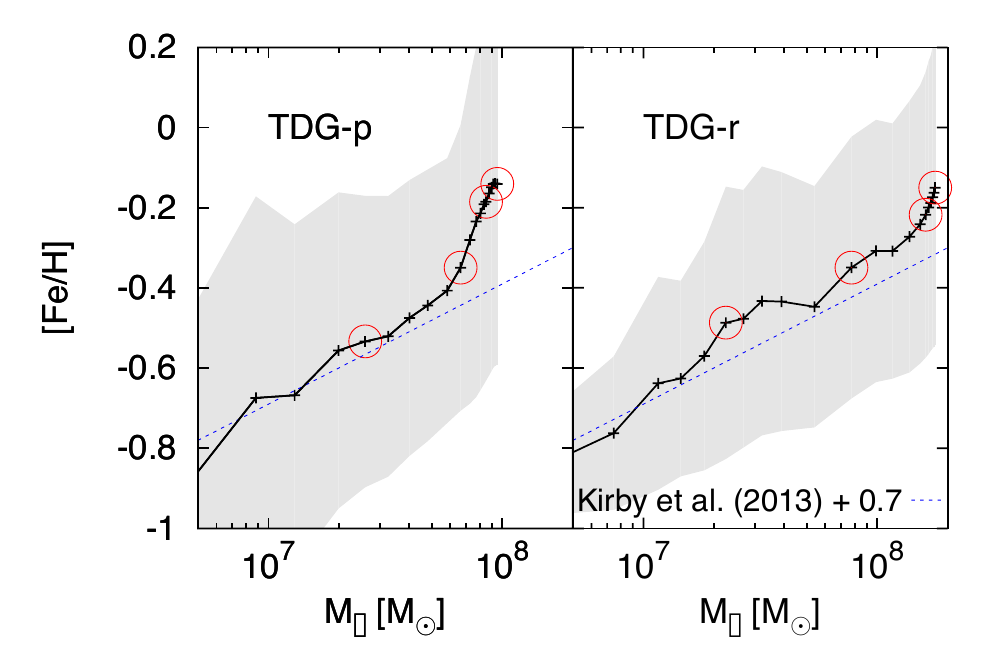}
		\caption{Iron enrichment of TDG-p (left panel) and \mbox{TDG-r} (right panel): For every 100 Myr, the mass-weighted average and standard deviation in [Fe/H] of all star clusters is calculated. The small crosses trace the evolution of the average [Fe/H] as the TDGs build up more stellar mass $M_{\star}$. The shaded area represents the $1\,\sigma$ region. The circles mark the position in this relation at 1.5, 2, 2.5, and 3 Gyr (from left to right). The dotted line is the MZ-relation (Eq.~\ref{eq:kirby}) from \citet{2013ApJ...779..102K} increased by 0.7 dex (see text).}
		\label{fig:MZrelation}
	\end{center}
\end{figure}

\noindent
Here, we start the simulations with significant pre-enrichment, typical for TDGs in the local Universe. Therefore an off-set between the \citet{2013ApJ...779..102K} data and the metallicity in the simulated TDGs is expected. Nevertheless, it is interesting to compare the increase in the iron abundance along the stellar mass growth to see if the TDGs evolve along a similar MZ-relation. Fig.~\ref{fig:MZrelation} shows the relation between the mass-weighted average in [Fe/H] and the stellar mass in 100 Myr steps for TDG-p and TDG-r. For illustration the 
observed MZ-relation (Eq.~\ref{eq:kirby}) increased by 0.7 ($[\mathrm{Fe/H}]=[\mathrm{Fe/H}]_{\mathrm {obs}}+0.7$) is shown. Especially TDG-r evolves very closely along the slope found in the LG DGs. This indicates that TDGs could build a similar MZ relation, with a normalisation that is dependent on the pre-enrichment and therefore the formation redshift (see Recchi, Kroupa \& Ploeckinger, subm.). A larger grid of initial TDG masses and metallicities will be presented in future publications.

In this work, we are interested in the maximum feedback case to investigate whether the simulated TDGs can survive the peri-centre passage and to study the dynamical effects of the tidal field on the TDG. For a more comprehensive study on the metal enrichment, the simplified model used here where every star particle represents a filled IMF, might be inaccurate. Truncated IMFs, as discussed in Paper I would reduce the ratio between SNII and SNIa and therefore lower the [O/Fe]. The values presented here, therefore serve as an upper limit. In addition, we do not include the mass flow along the tidal arms in our models.

We use the solar abundances from \citet{2009ARAA..47..481A} where $[\mathrm{O/Fe}]_{\odot} = 1.19$, $[\alpha \mathrm{/Fe}]_{\odot} = 1.25$, and  $[\mathrm{Fe/H}]_{\odot} = -4.5$.

\subsection{Cluster mass function}

\begin{figure}
\begin{center}
	\includegraphics[width=\linewidth]{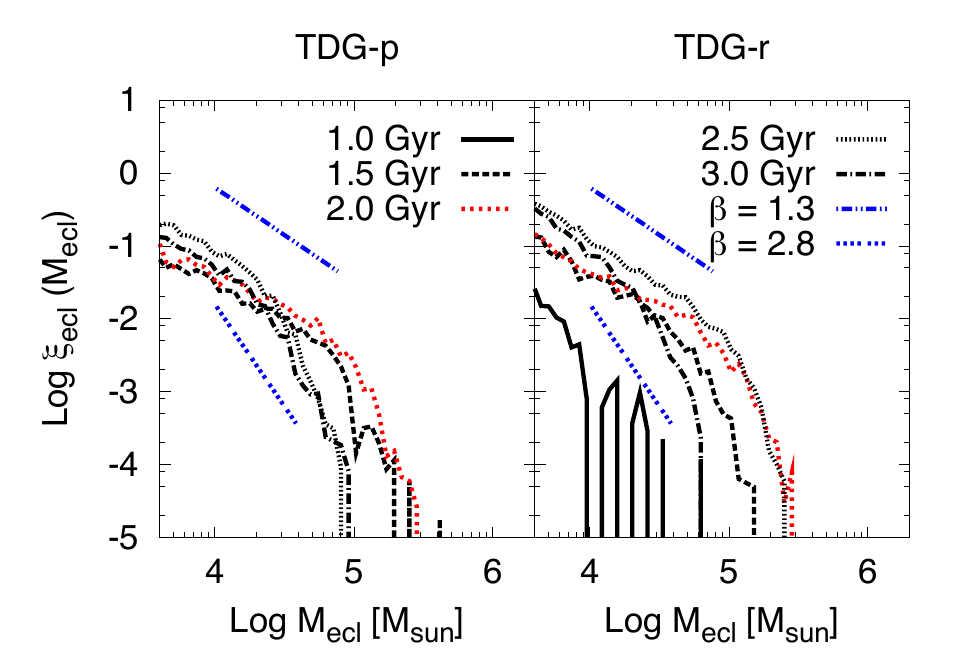}
	\caption{Star cluster mass function $\xi_{\mathrm{ecl}}$ for TDG-p (left) and TDG-r (right) at various times during the simulation. Note the shallower slope of $\xi_{\mathrm{ecl}}$ around the time of the peri-centre passage (t = 2 Gyr, highlighted in red in the color version). Slopes for $\beta = 1.3$ and $\beta = 2.8$ are indicated for guidance.}
\label{fig:ecmf}
\end{center}
\end{figure}

The distribution of embedded star cluster mass masses (embedded cluster mass function, ECMF) can be generally described by a Schechter distribution \citep{1976ApJ...203..297S} 
\begin{equation} \label{eq:schechter}
	\xi_{\mathrm{ecl}}(M) = \mathrm{d}N / \mathrm{d}M   \propto M^{-\beta} \, \mathrm{exp}(-M/M_c) \quad ,
\end{equation}

\noindent consisting of a power law with an index $\beta$ and an exponential truncation at $M_c$. Here dN is the number of clusters with stellar masses in the interval M to M + dM. As described in Sec.~\ref{sec:sfcriteria}, the mass of each star cluster particle depends on the local SF density and can span a wide range from less than $100\,\Msun$ to $10^6\,\Msunrm$ in our simulations. This allows to investigate the self-consistently formed distribution function of star cluster masses and a potential evolution of $\xi_{ecl}$ with time. In Fig.~\ref{fig:ecmf} we show the distribution of all star cluster masses that formed within the last 500 Myr at 1, 1.5, 2, 2.5, and 3 Gyr. At the low mass end ($M_{\mathrm{ecl}} < 10^4\,\Msun$) all distributions have slopes around $\beta \approx 1.3$, while for higher cluster masses ($M_{\mathrm{ecl}} > 10^4\,\Msun$) the slopes show a larger variance over the simulation time (Fig.~\ref{fig:ecmf}). The most shallow slopes are found at the peri-centre passage (Fig.~\ref{fig:ecmf}, t = 2 Gyr) when the compression by the tidal field is strongest and therefore more massive star clusters can be formed. Before and after the peri-centre passage, at t = 1 and 3 Gyr, the slopes are significantly steeper and in the transition to the exponential decay. In general, star clusters follow a slope of $\beta \approx 2$  \citep[e.~g.][]{2003ARAA..41...57L} but also shallower slopes can be found, as for example by \citet{2004MNRAS.347...17A} for a starburst DG (NGC 1569) with $\beta = 1.3$ to $1.75$. This agrees with our findings that the slopes are steeper during the self-regulated SF and shallower during the compression or starburst phase. TDG-t has a very low SFR ($\approx 10^{-4}\,\Msun \,\mathrm{yr}^{-1}$, see Fig.~\ref{fig:metallicity_new}) over its full evolution and therefore only star clusters up 130 $\Msun$ form in our description.

In the simulations presented in Paper I, the ECMF became top-heavy for a limited time during the collapse of the central part of the TDG. For the initial conditions presented here, where we already start with a rotationally supported TDG, we do not find a central collapse, but rather a general compression of the TDG by the tidal field. Therefore, the ECMF follows the general Schechter distribution. As only the local ISM determines the properties of each star cluster particle, the ECMF is modelled self-consistently with our approach and it can be used to investigate under which conditions the slopes of the Schechter distribution change or under which circumstances the ECMF can't be described with a Schechter distribution function.

\section{Discussion and conclusion} \label{sec:conclusion}

Based on the simulation setup reported in Paper I, we present in this work a long-term evolution study of TDGs. The main results can be summarised in the following categories:

\begin{enumerate}
	\item {\bf Tidal field:} The direct comparison between TDG-t (without a tidal field) and the simulation runs TDG-p and TDG-r (with a tidal field) demonstrates a drastic difference in the resulting properties. All three simulations start with mild SFRs of a few times $10^{-4}\,\Msun\,\mathrm{yr}^{-1}$, but while TDG-t stays at this low level and self-regulates its SF efficiently by stellar feedback, both TDGs in a tidal field experience a strong increase in the SFR after the apo-centre passage. As the TDGs approach the peri-centre, the tidal field compresses the TDG and the SF can remain at a high level without disrupting the TDG by stellar feedback. While gas is tidally stripped, especially in the first Gyr, when the TDG is more extended, and stellar feedback leads to smaller ejections perpendicular to the rotating disk, the TDG can re-accrete more material from the tidal arm, especially close to the peri-centre.
		
	\item {\bf Survivability:} The simulated TDGs in the maximum feedback case (with filled IMFs, compare: Paper I) survive for the complete simulation time of 3 Gyr. The chosen orbit is between typical radial distances of TDGs to the host galaxy, with a minimum analytical tidal radius of 4.3 kpc. More eccentric orbits with smaller tidal radii could lead to tidal disruption of the TDG, but this has not been demonstrated yet. In the simulations presented here, the compression by the tidal field and the stellar feedback balance without destroying the TDG. This is an indication that TDGs that form during galaxy--galaxy interactions over cosmic time may contribute significantly to the dwarf-galaxy satellite population. This would naturally explain the observed frequent occurrence of phase-space correlated satellite galaxy populations (see Sec.~\ref{sec:intro}).

	\item {\bf Rotation curves:} As the initial conditions represent a rotationally supported system, we investigate the influence of the tidal field on the rotation pattern of the TDG. Because in three TDGs flat rotation curves were observed by \citet{2007Sci...316.1166B} \citep[see also][]{2007AA...472L..25G}, we are especially interested whether the tidal field can disturb the dynamics of a TDG in such a way that the rotation curves flatten out. In the simulations here we do not find evidence for this process. As the resulting TDGs are more compact than the initial condition, their rotation curves decline even faster at later stages. As the TDGs in \citet{2007Sci...316.1166B} are young objects, we conclude that the observed flat rotation curves, can not be explained by a perturbing tidal field alone \citep[see][for further details]{2007AA...472L..25G}.

	\item {\bf Counter-rotation}  Throughout its evolution, the TDG experiences a torque from the variable tidal field, causing a pro-grade rotation. To study the effect of the initial rotation direction of the TDG on the long-term evolution, we set up a TDG with an initial pro-grade rotation (TDG-p) and a TDG that is initially rotating with a retro-grade spin (TDG-r).
We found counter-rotation of the ISM in TDG-r, most significantly after the peri-centre passage, but no clear signatures of counter-rotation are found in the stellar component. In TDG-p the gas is rotating in a pro-grade direction everywhere and a large fraction of the material that is tidally stripped, is lost. In comparison, in TDG-r the counter-rotation between the tidally stripped material and the centre of the TDG causes turbulence and more gaseous material is accreted back onto the TDG (compare t = 2.5 Gyr for TDG-p and TDG-r in Fig.~\ref{fig:gas}). Both simulations result in very compact objects with an extended period with high SFR, but as TDG-r accretes more material back from the tidal arm, it forms significantly more stars than TDG-p. 
Tidal interactions seem to be a promising process to create counter-rotating cores, without the need of merging two objects with opposite spin directions. Future work will reveal if after a full orbit, or for smaller peri-centric distances, also the stellar component can be affected and create a counter-rotating core, as observed in \citet{2014ApJ...783..120T} for dwarf elliptical galaxies in the Virgo cluster.

\item {\bf Metallicity}

The simulated TDGs start with a pre-enrichment of $Z = 0.1 \, \mathrm{Z}_{\odot}$ and solar abundance ratios. The initial metallicity is chosen to represent the outskirts of gas-rich galaxies in the local Universe to study the evolution of TDGs that have formed recently. The chemical evolution of each grid cell and each star cluster particle is traced throughout the simulation. We can reproduce general chemical properties of DGs but with an offset caused by the pre-enrichment. For example, the $\alpha$ abundances of all star clusters in the simulations show the typical ``knee" in the ([$\alpha$/Fe]-[Fe/H]) plot which is seen in LG DGs, but at higher values of [Fe/H]. Interestingly, the average [Fe/H] relative to the stellar mass $M_{\star}$ of the TDG evolves along a relation very similar to the MZ-relation found for LG DGs. While young TDGs always lie significantly above the MZ-relation for DGs \citep[e.~g.][]{1994AA...289...83D, 1998AA...333..813D, 2003AA...397..545W, 2009ApJ...705..723C, 2012AA...538A..61M}, TDGs that form out of less enriched gas could build a similar relation after several Gyr. We explore the evolution of the MZ-relation for TDGs in dependence of the pre-enrichment in terms of analytical chemical evolution models (Recchi, Kroupa \& Ploeckinger, subm.) and chemo-dynamical simulations (Paper III of this paper series) in separate works.

\end{enumerate}

\paragraph*{AGN feedback:}

A significant number of radio galaxies can be linked to galaxy interactions \citep[e.~g.][]{1986ApJ...306...64S, 1987ApJ...320..122H} and AGN (active galactic nuclei) feedback is an important driver for galaxy formation and evolution. TDGs could be affected by the gas outflow or the radiation caused by an AGN in the centre of the interacting galaxies. Assuming an interaction where the orbit as well as the galactic disks of the progenitor galaxies are in the same plane, any galactic outflow or AGN jet is expectedly perpendicular to the plane where the TDGs form. For more general interactions with arbitrary inclinations, a potential galactic outflow could overlap with a tidal arm and influence the formation or evolution of TDGs. \citet{2014MNRAS.437.2137S} studied close galaxy pairs and did not find a large difference in the Seyfert fraction when compared to isolated galaxies. They conclude that the AGN phase might become dominant in the very final stages of the merger process once the black holes have coalesced, in agreement with similar conclusions by \citet{2012MNRAS.420.2139C}. \citet{2006ApJ...643..692C} simulated galaxy mergers with and without black holes and found clear differences in the X-ray halo
of the completely merged galaxies, especially in the innermost 50 kpc. At earlier times, when the tidal arms are prominent and at larger distances to the mass centre, the influence of the BH feedback on the temperature and density of the halo gas is negligible. We focus here on the first 3 Gyr after the first encounter and on distances between the TDG and the mass centre of the interacting galaxies of 64 to 135 kpc. While AGN feedback undoubtedly has a strong impact on the energy and metallicity of galactic haloes, we neglect it for our simulations as it would influence the evolution of TDGs only in a narrow range of orbital parameter and TDG formation times.

\section{Acknowledgements}
We want to thank the referee for carefully reading the manuscript and very appreciated comments that helped to improve and clarify the original version of this paper. The authors are grateful for simulation data on galaxy interactions provided by Francois Hammer and Yanbin Yang that provide a very helpful guidance for the initial conditions used in the chemo-dynamical simulations. 
The software used in this work was in part developed by the DOE-supported ASC / Alliance Center for Astrophysical Thermonuclear Flashes at the University of Chicago. The numerical simulations are performed at the HPC astro-cluster of the Institute of Astronomy and at the Vienna Scientific Cluster (VSC1) under project no. 70128.

\bibliographystyle{mn2e}
\bibliography{../../PHD_THESIS_comb/Ploeckinger_thesis}

\section{Appendix} \label{sec:appendix}

In this paper we perform numerical experiments of DM-free DGs, to investigate the influence of a tidal field on their properties and stability. The initial conditions of our setup represent properties that are found in large-scale galaxy interaction simulations. 
We analysed the TDGs that form in the simulation of an ancient merger at M31 \citep[see][]{2010ApJ...725..542H}
and for which the tidal tail \citep[see][]{2012MNRAS.427.1769F, 2014MNRAS.442.2419Y} has been retrieved at a very early
stage, 1.5 Gyr after the first encounter of the progenitor galaxies.

\begin{figure}
	\includegraphics*[width = \linewidth] {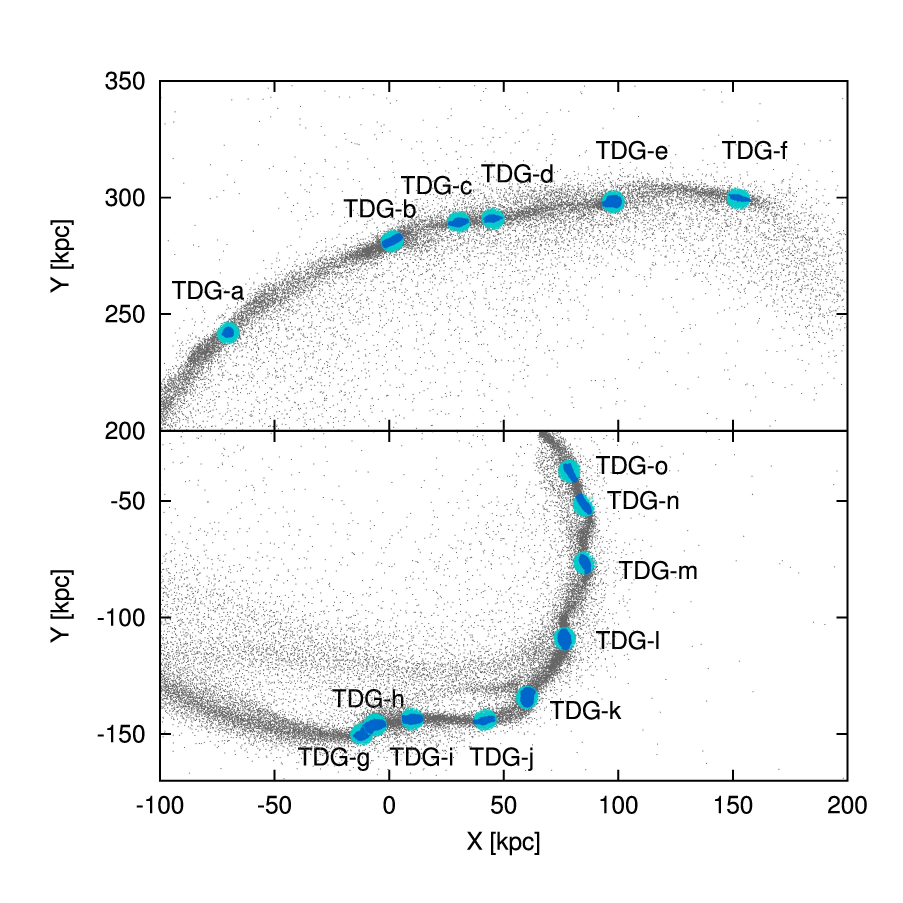}
	\caption{Simulation snapshot from \citet{2012MNRAS.427.1769F} at 1.5 Gyr after the first encounter of the interacting galaxies. The two panels show the two main tidal arms at this time. The grey dots indicate the positions of the tidal arm. For illustration, only every tenth SPH particle is plotted. The circles have a radius of $r_{\mathrm{TDG}} = 5 \,\mathrm{kpc}$ and mark the positions of the identified proto-TDGs. The high-density ($\rho \ge 5 \times 10^{-26} \,\mathrm{g\,cm}^{-3}$) SPH particles within the proto-TDGs are highlighted within each circle.}
	\label{fig:pos_hammer}
\end{figure}

We identify 15 forming TDGs in the two most prominent tidal arms at this snapshot (Fig.~\ref{fig:pos_hammer}). For each TDG, we study its properties within a radius of $r_{\mathrm{TDG}} = 5\,\mathrm{kpc}$. The value is chosen to match the extent of all identified objects.
For all 15 TDGs, the following properties are derived:

\paragraph*{Mass centre $\vec{R}_{\mathrm{M}}$ :} Based on their approximate position, the mass centre of all SPH (smoothed particle hydrodynamics) particles with masses $m_i$ and positions $\vec{r}_i$  within $r_{\mathrm{TDG}}$ is determined by

\begin{equation}
\vec{R}_{\mathrm{M}} = \frac{1}{\sum_i m_i} \sum_i m_i \vec{r}_i \;.
\end{equation}

\noindent
 As the new mass centre is the origin for a different 5 kpc sphere, $\vec{R}_{\mathrm{M}}$  is calculated iteratively until convergence is reached. 

\paragraph*{Orbit velocity $\vec{v}_{\mathrm{orbit}}$:} The mass weighted average velocity of particles within $r_{\mathrm{TDG}}$ is calculated. This describes the system velocity of each TDG and determines the orbit around the main galaxies. 

\begin{figure}
	\includegraphics*[width = 0.7\linewidth] {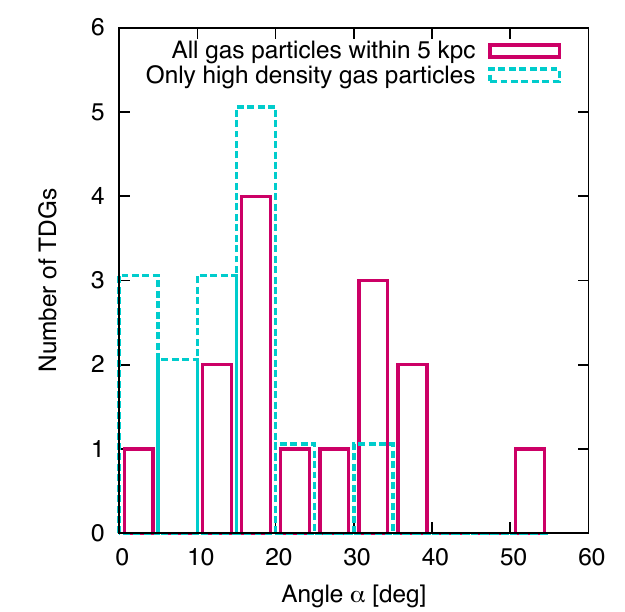}
	\caption{Number of identified proto-TDGs from the snapshot of \citet{2012MNRAS.427.1769F} in bins of $\Delta \alpha = 5 \, \deg$. The red, solid boxes show the distribution if all SPH particles within the r = 5 kpc sphere are considered for the calculation, while the green, dashed boxes show the distribution if only the high density ($\rho \ge 5 \times 10^{-26} \,\mathrm{g\,cm}^{-3}$)  SPH particles are considered.}
	\label{fig:angle_hammer}
\end{figure}

\paragraph*{Orbital angular momentum $\vec{L}_{\mathrm{orbit}}$:} The angular momentum vector of the trajectory of each TDG around the main galaxies is calculated with
 
\begin{equation}
	\vec{L}_{\mathrm{orbit}} = \vec{R}_{\mathrm{M}} \times (M_{\mathrm{TDG}} \cdot \vec{v}_{\mathrm{orbit}}) \, ,
\end{equation}

\noindent
where $M_{\mathrm{TDG}}$ is the gas mass within $r_{\mathrm{TDG}}$.

\paragraph*{Angular momentum of each TDG $\vec{L}_{\mathrm{TDG}}$:} In oder to get the internal velocity pattern of each TDG, the rest-frame of the TDG is used for further analysis. Therefore, the velocity $\vec{v}_i$ of each particle with mass $m_i$ is reduced by the average velocity $\vec{v}_{\mathrm{orbit}}$ and the position of each particle $\vec{p}_i$ is reduced by $\vec{R}_{\mathrm{M}}$. In the rest-frame of the TDG, the angular momentum of the rotation within the TDG is calculated as the sum over all particles $i$

\begin{equation}
	\vec{L}_{\mathrm{TDG}} = \sum_i (\vec{p}_i - \vec{R}_{\mathrm{M}}) \times [m_i \cdot (\vec{v}_i - \vec{v}_{\mathrm{orbit}})] \, .
\end{equation}

\paragraph*{Angle $\alpha$ between $\vec{L}_{\mathrm{orbit}}$ and $\vec{L}_{\mathrm{TDG}}$:} 
The angle $\alpha$ gives information about the spin direction of the TDG relative to the angular momentum direction of the orbital motion. For $\alpha < 90\,\deg$ the rotation of the TDG is pro-grade relative to its orbit, while an angle of $\alpha > 90\,\deg$ indicates a retro-grade rotation. In Fig.~\ref{fig:angle_hammer} a histogram of the distribution of $\alpha$ for the identified TDGs is plotted. All TDGs analysed here show a pro-grade rotation with even smaller angles $\alpha$ for the high density ($\rho \ge 5 \times 10^{-26} \,\mathrm{g\,cm}^{-3}$) SPH particles. In all cases, the rotation direction is well correlated with their orbit. A narrowly aligned rotation would be expected if the internal rotation is caused by the tidal field.

\paragraph*{Rotation curves:} For each TDG we construct rotation curves from the reduced velocities and positions. As the main rotation plane is defined by its normal vector $\vec{L}_{\mathrm{TDG}}$, the tangential velocity component for each particle $i$ points towards the direction of $\hat{v}_{i,T} = \hat{L}_{\mathrm{TDG}} \times \hat{r}_i$ with $ \hat{L}_{\mathrm{TDG}} = \vec{L}_{\mathrm{TDG}} / |\vec{L}_{\mathrm{TDG}}|$ and $\hat{r}_i =(\vec{p}_i - \vec{R}_{\mathrm{M}}) /  |\vec{p}_i - \vec{R}_{\mathrm{M}} |$. 
The absolute value of the rotation velocity of each particle is the projection onto the tangential direction:

\begin{equation}
	v_{\mathrm{i,rot}} =  | (\vec{v}_i - \vec{v}_{\mathrm{orbit}}) \cdot \hat{v}_{i,T} |
\end{equation}

\begin{figure}
	\includegraphics*[width = \linewidth] {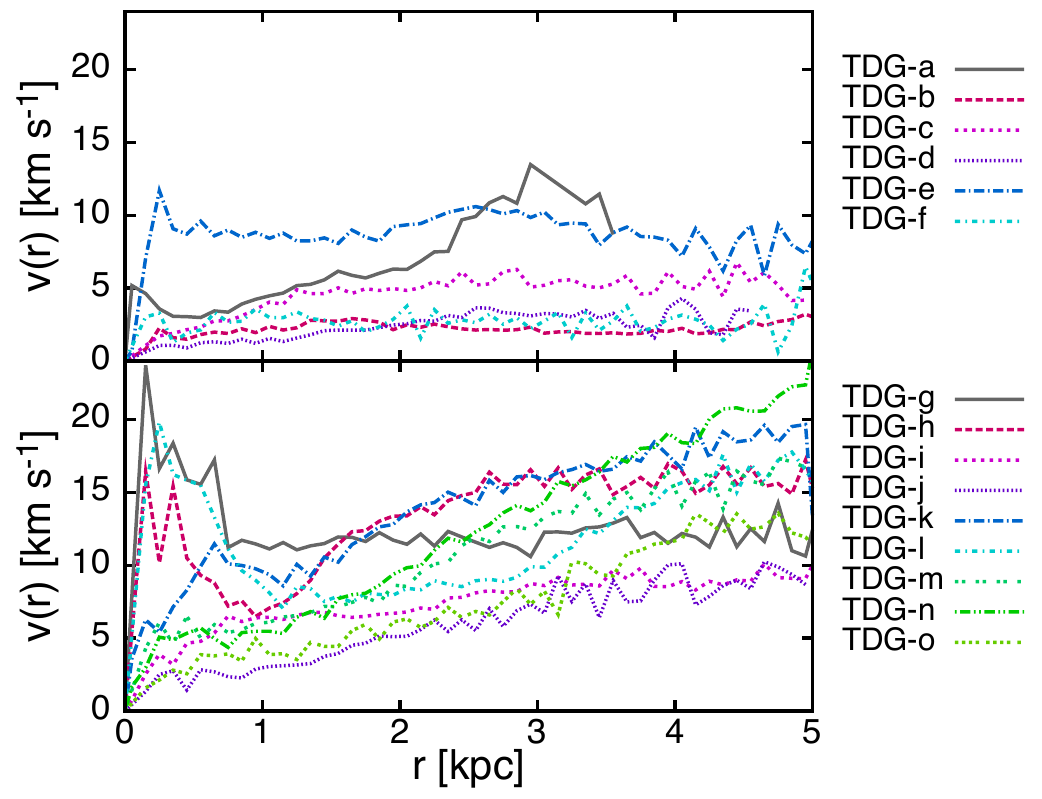}
	\caption{Rotation curves for all identified proto-TDGs from the snapshot of \citet{2012MNRAS.427.1769F}. The top panel shows the rotation curves for TDGs in the Y $>$ 0 tidal arm (top panel of Fig.~\ref{fig:pos_hammer}) and the bottom panel shows the rotation curves for TDGs in the Y $<$ 0 tidal arm (bottom panel of Fig.~\ref{fig:pos_hammer}).}
	\label{fig:vrot_hammer}
\end{figure}

\noindent
For the construction of the rotation curves (Fig.~\ref{fig:vrot_hammer}) all $v_{\mathrm{i,rot}}$ are distributed in radial distance bins with $\Delta r = 0.1\,\mathrm{kpc}$ and for each radial bin, the mass weighted average is calculated. As seen in Fig.~\ref{fig:vrot_hammer} not all TDGs have already a clear rotation signature at this early stage. TDGs with a gradually rising rotation curve might not be kinematically decoupled from the tidal arm. Mildly increasing rotation curves are seen in our simulations at t = 500 Myr during an initial expansion period (Fig.~\ref{fig:rot_all}).

\paragraph*{Velocity dispersion $\sigma_{\mathrm{TDG}}$:} In order to study whether a TDG is rotationally or pressure supported, we calculate the velocity dispersion for each TDG with:

\begin{equation}
	\sigma_{\mathrm{TDG}}^2 =  \frac{1}{M_{\mathrm{TDG}}} \sum_i (| \vec{v}_i  - \vec{v}_{\mathrm{orbit}}| )^2
\end{equation}

\noindent
Table~\ref{tab:hammerTDGs_low} summarises the derived values for the above mentioned quantities for all particles within $r_{\mathrm{TDG}}$ and Table~\ref{tab:hammerTDGs_high} presents the properties when only the high density particles ($\rho \ge 5 \times 10^{-26} \,\mathrm{g\,cm}^{-3}$) are taken into account.

\begin{table*}
\caption{This table lists the properties of the analysed TDGs from a galaxy interaction simulation of \citet{2012MNRAS.427.1769F}. For this table all gas particles within a 5 kpc sphere are considered to calculate the TDG quantities, while in Table~\ref{tab:hammerTDGs_high} only high density particles are considered. Column {\bf 1}: TDG identifier as indicated in Fig.~\ref{fig:pos_hammer}. Column {\bf 2-4}: Position (X, Y, Z) of the mass centre of the TDG relative to the mass centre of the interacting galaxies. Column {\bf 5}: Total mass of all gas particles within a 5 kpc sphere around $\vec{R}_{\mathrm{M}}$. Column {\bf 6}: Angle between the rotation directions of the TDG orbit and the internal TDG rotation. Column {\bf 7-9}: Orbit velocity $(v_x, v_y, v_z)$ of the TDGs. Column {\bf 10}: Velocity dispersion within the TDG. }
\begin{center}
\begin{tabular}{c|ccc|c|ccc|c|c|c}
\hline
\hline
TDG		&	\multicolumn{3}{c}{$\vec{R}_{\mathrm{M}}$}	& 	$M_{\mathrm{TDG}}$	& $\alpha$   	& \multicolumn{3}{c}{$\vec{v}_{\mathrm{orbit}}$ }		& $\sigma$ 	 \\
		&	\multicolumn{3}{c}{$[\mathrm{kpc}]$}			&	$[10^8\,\Msun]$		& $[\deg]$    	&\multicolumn{3}{c}{$[\mathrm{km\,s}^{-1}]$}			& $[\mathrm{km\,s}^{-1}]$ \\
\hline

a  &   -70.27   &   242.12   &   59.83   &   2.07  &   30.91   &   -94.55  &   117.44   &   10.41  &   6.65  \\  
b  &   1.39   &   281.37   &   102.25   &   2.15  &   54.53   &   -54.06  &   161.56   &   51.65  &   5.73  \\  
c  &   30.31   &   289.39   &   114.06   &   1.87  &   35.02   &   -35.15  &   171.82   &   64.81  &   6.20  \\  
d  &   45.36   &   290.94   &   119.38   &   1.24  &   26.59   &   -27.71  &   177.38   &   71.70  &   5.93  \\  
e  &   97.78   &   298.18   &   128.32   &   3.35  &   30.18   &   7.13  &   188.94   &   88.80  &   9.80  \\  
f  &   152.31   &   299.67   &   122.14   &   0.97  &   20.90   &   43.87  &   198.29   &   94.78  &   7.55  \\  
g  &   -11.88   &   -150.09   &   105.53   &   2.88  &   38.25   &   82.30  &   -86.11   &   43.35  &   13.39  \\  
h  &   -6.16   &   -146.04   &   100.64   &   3.61  &   11.02   &   81.03  &   -71.27   &   39.25  &   13.43  \\  
i  &   10.16   &   -143.58   &   91.29   &   2.39  &   17.32   &   86.25  &   -59.51   &   23.97  &   6.74  \\  
j  &   41.93   &   -143.90   &   81.18   &   1.43  &   33.71   &   105.96  &   -46.36   &   3.26  &   6.11  \\  
k  &   60.31   &   -134.38   &   69.63   &   5.47  &   19.18   &   105.65  &   -28.01   &   -12.48  &   15.27  \\  
l  &   76.55   &   -109.19   &   42.70   &   4.44  &   2.71   &   92.09  &   13.01   &   -44.19  &   9.39  \\  
m  &   85.37   &   -76.59   &   10.43   &   2.64  &   18.35   &   59.45  &   66.79   &   -76.27  &   8.37  \\  
n  &   85.13   &   -51.73   &   -11.02   &   2.30  &   14.48   &   14.09  &   103.98   &   -84.95  &   9.43  \\  
o  &   78.74   &   -37.27   &   -20.37   &   1.45  &   17.93   &   -26.31  &   116.96   &   -75.68  &   8.07  \\

\hline
\end{tabular}
\end{center}
\label{tab:hammerTDGs_low}
\end{table*}%

\begin{table*}
\caption{Columns {\bf 1-10} as in Table~\ref{tab:hammerTDGs_low} but here only the high density ($\rho \ge 5 \times 10^{-26} \,\mathrm{g\,cm}^{-3}$) gas particles are considered. Column {\bf 11, 12}: For all TDGs with a clear rotation curve, the maximum rotation velocity $v_{\mathrm{r,max}}$, as well as the radius where $v(r) = v_{\mathrm{r,max}}$ is listed. }
\begin{center}
\begin{tabular}{cccccccccccccc}
\hline
\hline
TDG		&	\multicolumn{3}{c}{$\vec{R}_{\mathrm{M}}$}	& 	$M_{\mathrm{TDG}}$	& $\alpha$  	& \multicolumn{3}{c}{$\vec{v}_{\mathrm{orbit}}$ }	& $\sigma$ 			& $v_{\mathrm{r,max}}$  		&  $r(v_{\mathrm{r,max}})$\\
		&	\multicolumn{3}{c}{$[\mathrm{km\,s}^{-1}]$}	&	$[10^8\,\Msun]$		& $[\deg]$		&\multicolumn{3}{c}{$[\mathrm{km\,s}^{-1}]$}		& $[\mathrm{km\,s}^{-1}]$	& $[\mathrm{km\,s}^{-1}]$ 		& [kpc]\\
\hline
a  &   -70.27   &   242.12   &   59.83   &   2.07  &   30.91   &   -94.55  &   117.44   &   10.41  &   6.65  &  		5.19 		&	0.1 	\\  
b  &   1.39   &   281.37   &   102.25   &   2.15  &   54.53   &   -54.06  &   161.56   &   51.65  &   5.73  &   		2.91 		&	1.7	\\  
c  &   30.31   &   289.39   &   114.06   &   1.87  &   35.02   &   -35.15  &   171.82   &   64.81  &   6.20  &		-		&	- 	\\  
d  &   45.36   &   290.94   &   119.38   &   1.24  &   26.59   &   -27.71  &   177.38   &   71.70  &   5.93  & 		-		&	-	\\  
e  &   97.78   &   298.18   &   128.32   &   3.35  &   30.18   &   7.13  &   188.94   &   88.80  &   9.80  &       		11.70	&	0.3 \\  
f  &   152.31   &   299.67   &   122.14   &   0.97  &   20.90   &   43.87  &   198.29   &   94.78  &   7.55  &    		-		&	-	\\  
g  &   -11.88   &   -150.09   &   105.53   &   2.88  &   38.25   &   82.30  &   -86.11   &   43.35  &   13.39  &      	23.73	&	0.2 \\  
h  &   -6.16   &   -146.04   &   100.64   &   3.61  &   11.02   &   81.03  &   -71.27   &   39.25  &   13.43  &    		16.42	&	0.2\\  
i  &   10.16   &   -143.58   &   91.29   &   2.39  &   17.32   &   86.25  &   -59.51   &   23.97  &   6.74  &    		-		&	- \\  
j  &   41.93   &   -143.90   &   81.18   &   1.43  &   33.71   &   105.96  &   -46.36   &   3.26  &   6.11  &    		-		&	- \\  
k  &   60.31   &   -134.38   &   69.63   &   5.47  &   19.18   &   105.65  &   -28.01   &   -12.48  &   15.27  &   	11.46	&	0.7 \\  
l  &   76.55   &   -109.19   &   42.70   &   4.44  &   2.71   &   92.09  &   13.01   &   -44.19  &   9.39  &    		19.82	&	0.3\\  
m  &   85.37   &   -76.59   &   10.43   &   2.64  &   18.35   &   59.45  &   66.79   &   -76.27  &   8.37  &    		-		&	-\\  
n  &   85.13   &   -51.73   &   -11.02   &   2.30  &   14.48   &   14.09  &   103.98   &   -84.95  &   9.43  &    		-		&	-\\  
o  &   78.74   &   -37.27   &   -20.37   &   1.45  &   17.93   &   -26.31  &   116.96   &   -75.68  &   8.07  &    		-		&	-\\ 
\hline
\end{tabular}
\end{center}
\label{tab:hammerTDGs_high}
\end{table*}%

\end{document}